\documentclass[pdftex,twocolumn,epjc3]{svjour3}          

\RequirePackage[T1]{fontenc}

\RequirePackage{graphicx}
\RequirePackage{subfigure}
\RequirePackage{epstopdf}
\RequirePackage{tabularx}

\RequirePackage{feynmp}
 \RequirePackage[utf8]{inputenc}

\RequirePackage{float}

\journalname{Eur. Phys. J. C}

\def\vecptmiss{\ensuremath{\vec{p}_T^{\mathrm{\, miss}}}}
\def\ptmiss{\ensuremath{p_T^{\mathrm{\, miss}}}}
\def\etmiss{\ensuremath{E_T^{\mathrm{\, miss}}}}

\def\runIbound{1080 GeV}
\def\runIIbound{1900 GeV}

\begin{document}


\title{Prompt Signals and Displaced Vertices 
  in Sparticle Searches for Next-to-Minimal
  Gauge Mediated Supersymmetric Models} 

\author{B.~C.~Allanach \thanksref{a, addr1} \and
Marcin Badziak \thanksref{b, addr2, addr3, addr4} \and
Giovanna Cottin \thanksref{c, addr5} \and
Nishita Desai \thanksref{d, addr6} \and
Cyril Hugonie \thanksref{e, addr7} \and
Robert Ziegler \thanksref{f, addr8, addr9}}
\thankstext{a}{b.c.allanach@damtp.cam.ac.uk}
\thankstext{b}{mbadziak@fuw.edu.pl}
\thankstext{c}{gfc24@cam.ac.uk}
\thankstext{d}{n.desai@thphys.uni-heidelberg.de}
\thankstext{e}{cyril.hugonie@umontpellier.fr}
\thankstext{f}{robert.ziegler@lpthe.jussieu.fr}

\institute{\label{addr1}Department of Applied Mathematics and Theoretical Physics, Centre for Mathematical Sciences, University of Cambridge, Wilberforce Road, Cambridge CB3 0WA, United Kingdom \and
\label{addr2} Institute of Theoretical Physics, Faculty of Physics,
  University of Warsaw, ul.\ Pasteura 5, PL-02-093 Warsaw, Poland \and
\label{addr3} Department of Physics, University of California, Berkeley, CA 94720, USA \and
\label{addr4} Ernest Orlando Lawrence Berkeley National Laboratory,
University of California, Berkeley, CA 94720, USA \and
\label{addr5} Cavendish Laboratory, University of Cambridge, 19 JJ Thomson Ave, Cambridge CB3 0HW,
  United Kingdom \and
\label{addr6} Institut f\"{u}r Theoretische Physik, Philosophenweg 16, 69120, Heidelberg, Germany \and
\label{addr7} LUPM, UMR 5299, CNRS, Universit\'{e} de Montpellier, 34095, Montpellier, France \and
\label{addr8} Sorbonne Universit\'{e}s, UPMC Univ Paris 06, UMR 7589, LPTHE, F-75005, Paris, France \and
\label{addr9} CNRS, UMR 7589, LPTHE, F-75005, Paris, France}

\maketitle


\begin{abstract}We study the LHC phenomenology of the next-to-minimal model of gauge-mediated supersymmetry breaking (NMGMSB), both for Run I and Run II. The Higgs phenomenology of the model is consistent with observations: a 125 GeV Standard Model-like Higgs which mixes with singlet-like state of mass around 90 GeV that provides a 2$\sigma$ excess at LEP II. The model possesses regions of parameter space where a longer-lived lightest neutralino decays in the detector into a gravitino and a $b-$jet pair or a tau pair resulting in potential displaced vertex signatures. We investigate current bounds on sparticle masses and the discovery potential of the model, both via conventional searches and via searches for displaced vertices. The searches based on promptly decaying sparticles currently give a lower limit on the gluino mass 1080 GeV and could be sensitive up to 1900 GeV with 100~fb$^{-1}$, whereas the current displaced vertex searches cannot probe this model due to $b-$quarks in the final state. We show how the displaced vertex cuts might be relaxed in order to improve signal efficiency, while simultaneously applied prompt cuts reduce background, resulting in a much better sensitivity than either strategy alone and motivating a fully fledged experimental study.
  \end{abstract}

\section{Introduction}

Sparticle searches at the Large Hadron Collider (LHC) have so far
yielded no clear discovery. Strengthening exclusion
limits~\cite{Aad:2015iea} on the masses of sparticles in
the minimal supersymmetric standard model (MSSM) mean that points in
the model parameter space with low fine tuning have been ruled
out. In particular, the ATLAS and CMS experiments have measured a
particle whose properties are compatible with a Standard Model (SM)
Higgs of mass around 125 GeV. While such a mass is still compatible
with the theoretical upper bound in the MSSM, it is rather on the
heavy side and corresponds to a fairly heavy stop mass, which in turn
induces lower bounds on typical quantifications of fine-tuning in the
electroweak symmetry breaking sector (see e.g. Ref.~\cite{Hall:2011aa}).  Arguably, it may be important
to consider non-minimal supersymmetric scenarios that can alter the
interpretation of standard sparticle searches, perhaps allowing regions
with lower fine tuning than in minimal scenarios.  However, one may get an
impression from 
``simplified model searches'' (where the MSSM spectrum is set to be
heavy except for a few sparticles relevant for a particular search)
that strongly interacting particles with multi-TeV masses are already
ruled out (see, for example Ref.~\cite{Aad:2015iea}), eliminating the
low fine-tuning regions. Still, in more realistic non-simplified
MSSM scenarios, points in parameter space exist with gluino masses of
700 GeV or squark masses of 500 GeV which evade all Run I sparticle
searches~\cite{Aad:2015baa} or even which evade all 2015 Run II
searches~\cite{Barr:2016inz} with consequently fairly low values of
fine-tuning.

Non-simplified MSSM scenarios are suggested by well motivated ultra-violet
 scenarios of SUSY breaking, for example Gauge Mediation.  While gauge 
mediated SUSY breaking provides a neat solution to the SUSY flavour
problem (i.e.\ the absence of large sources of flavor violation in the soft terms),
its minimal realisations are in trouble because they typically predict a SM-like
Higgs mass that is too low compared to the observed value around 125
GeV. A potentially fruitful path was explored by introducing
additional dynamics to increase the SM-like Higgs boson mass
prediction while maintaining fairly low levels of fine-tuning, see Refs.~\cite{Evans:2011bea,Jelinski:2011xe,Evans:2012hg,Kang:2012ra,Craig:2012xp,Abdullah:2012tq,Kim:2012vz,Byakti:2013ti,Craig:2013wga,Evans:2013kxa,Calibbi:2013mka,Jelinski:2013kta,Galon:2013jba,Fischler:2013tva,Knapen:2013zla,Ding:2013pya,Calibbi:2014yha,Basirnia:2015vga,Jelinski:2015gsa,Jelinski:2015voa,Knapen:2015qba}.
 
In Ref.~\cite{Allanach:2015cia}, we revisited a simple model by
Delgado, Giudice and Slavich~\cite{Delgado:2007rz} (DGS) that combines
gauge mediation (GM) and the next-to-minimal supersymmetric Standard
Model (NMSSM).  The field content of the model is the one of the NMSSM, plus two copies of messengers in $\bf 5 +\bar 5$
representations of SU(5), denoted by $\Phi_i, \bar \Phi_i$,
respectively ($i \in \{ 1,\ 2\}$), with doublet and triplet components
$\Phi^D_i, \bar \Phi_i^D$ and $\Phi^T_i, \bar \Phi_i^T$. SUSY breaking is parameterised by the spurion $X= M + F
\theta^2$ (where $M$ is the messenger scale and $\theta$ is the
Grassmann valued $N=1$ superspace coordinate). Aside from Yukawa
interactions, the superpotential contains spurion-mes\-sen\-ger couplings and singlet $S$-messenger couplings (first introduced in 
the context of gauge mediation in Ref.~\cite{Giudice:1997ni})
\begin{eqnarray}
W & = & \ldots + \lambda S H_u H_d + \frac{\kappa}{3}S^3 \nonumber \\
& & + X \sum_i (\kappa_i^D \bar \Phi_i^D \Phi_i^D + \kappa_i^T \bar \Phi_i^T\bar \Phi_i^T) \nonumber \\ 
& & + S(\xi_D \bar \Phi_1^D \Phi_2^D + \xi_T \bar \Phi_1^T \Phi_2^T),
\end{eqnarray}
where the singlet-messenger couplings unify at the grand
unified theory scale $M_{\rm GUT}$: $\xi_D(M_{\rm GUT})=\xi_T(M_{\rm GUT}) \equiv \xi $  with unified coupling $\xi$. The scale
of the SUSY breaking terms is fixed by the parameter $\tilde m = 1/(16
\pi^2) F/M$.

\begin{figure}[ht]
\centering \includegraphics[width=\columnwidth]{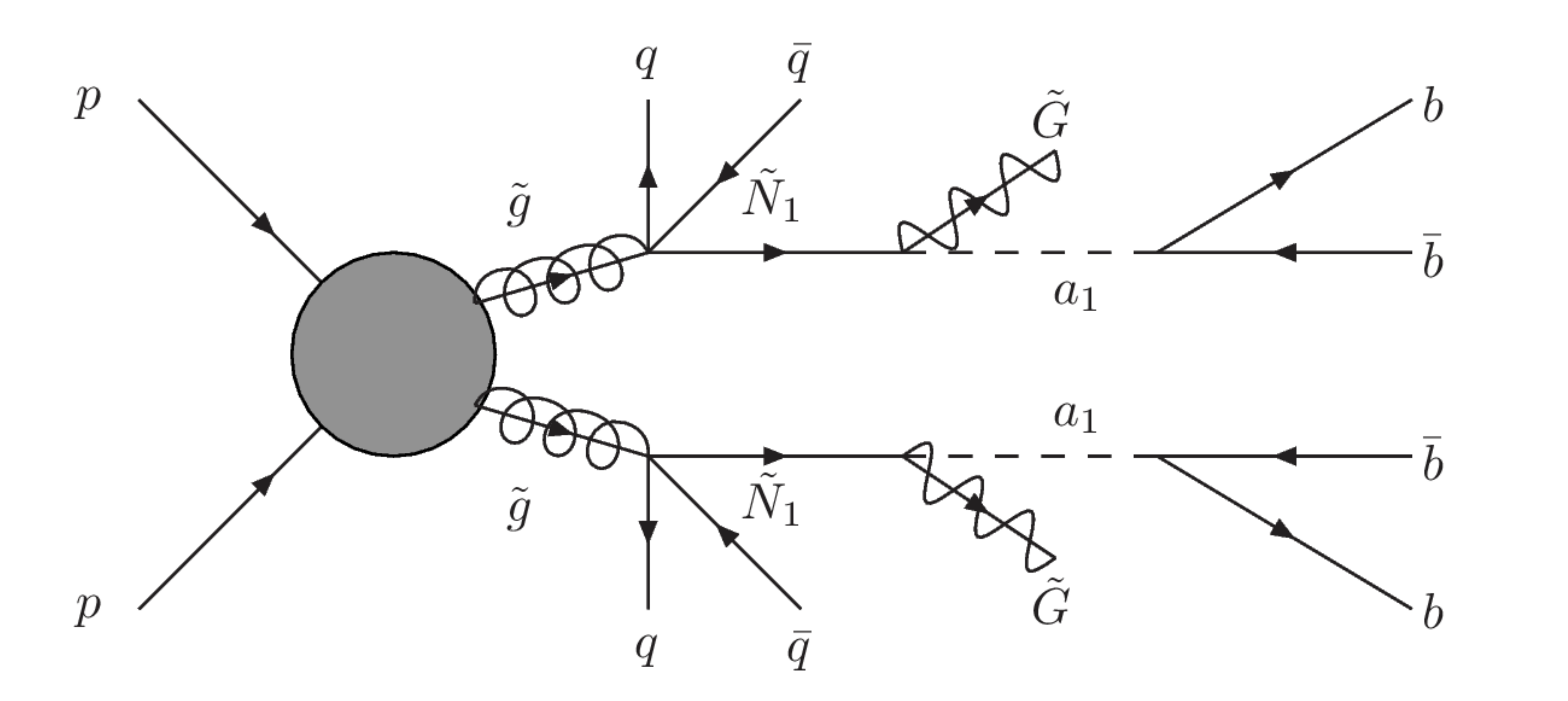}
\caption{An example of LHC sparticle production in the DGS model, followed by sparticle decay. In this example, we have
  four hard prompt jets from gluinos decaying into quarks $q$ and
  anti-quarks $\bar q$; the lightest neutralino $\tilde{N}_1$ may have
  an intermediate life-time, producing displaced vertices, each generating
  $b \bar b$. The gravitino $\tilde G$ leaves a missing transverse
  momentum signature. The lightest pseudo-scalar $a_1$ has a lower
  branching ratio for decays into $\tau \bar \tau$ than $b \bar b$.
  The $\tilde g \rightarrow \tilde N_1$ part of the decay may commonly
  be more complicated, involving a cascade decay and concomitant
  additional SM states.\label{fig:feyn}}
\end{figure}

It was shown in Ref.~\cite{Allanach:2015cia} that in the DGS model one can obtain a 125 GeV Standard Model-like Higgs boson with stops as
light as 1.1 TeV, thanks to the mixing of the Higgs with a singlet
state at ${\cal O} (90-100)$ GeV which is compatible with LEP
data~\cite{Badziak:2013bda}. With these Higgs constraints, essentially
all parameters are fixed except for the GM messenger scale which
mainly controls the phenomenology of the gravitino. The central
feature of the model, apart from the light Higgs that
might explain the LEP excess~\cite{Schael:2006cr} is the peculiar structure of
the light 
sparticle spectrum.  The lightest sparticle (LSP) is the
gravitino\footnote{Another attractive feature of the NMSSM realisation of gauge
  mediation is that the singlet allows the gravitino to be a good dark
  matter candidate even for large reheating temperatures that are
  compatible with thermal leptogenesis~\cite{Badziak:2015dyc}.} $\tilde G$
with
mass and couplings effectively set by the GM messenger scale, the
next-to-LSP (NLSP) is a singlino-like neutralino $\tilde{N}_1$ of mass
around 100 GeV, and the next-to-NLSP (NNLSP) is a bino-like neutralino
$\tilde{N}_2$ or stau $\tilde \tau$, depending on the GM messenger
scale. The presence of the singlino alters SUSY decay chains as
compared to the MSSM, leading to additional $b$-jets or taus.
One distinctive feature of this scenario is that the singlino
decays to a gravitino and a light singlet-like pseudoscalar $a_1$ of mass around 20 GeV, with the
latter decaying predominantly to $b\bar{b}$ as well as to $\tau\tau$.  Depending on the GM
messenger scale, the two $b$-jets may be produced far outside the
detector (when the $\tilde{N}_1$ is quasi-stable, at high GM scales)
or at low GM scales, they may be produced within the detector from displaced
vertices (DVs). This peculiar feature of a long-lived singlino decay was already 
noticed in Ref.~\cite{Delgado:2007rz}.  An
example diagram showing LHC sparticle production in the model is shown
in Figure~\ref{fig:feyn}.

In this paper, we wish to evaluate the collider phenomenology of the
model. In section~\ref{sec:bench}, we describe our benchmark model, and describe 
the tools used for simulation of the signal events and validation of our analysis.  In 
section~\ref{sec:prompt}, we re-cast the most constraining
prompt sparticle searches from the LHC in order to find out how
stringent the bounds on the model are and then we estimate the future reach. 
In section~\ref{sec:displaced}, we detail a study of DV signatures, 
starting with recasting the current ATLAS multi-track DV + jets analysis and showing that 
current searches are not sensitive to our model. By changing the cuts, we
suggest ways in which the DV cuts can be loosened, and
how cuts on accompanying hard prompt objects can be used to combat
background rates. In section~\ref{sec:DVrunii}, we estimate the search
reach from early Run II data with such a strategy. After a summary and
discussion in section~\ref{sec:summary}, we define the relevant
DV variables $d_0$ and $r_{DV}$ in
Appendix~\ref{sec:def}, for easy reference by the reader.

\section{Benchmark model and event generation \label{sec:bench}}

\begin{figure*}[ht]
\begin{center}
\includegraphics[width=0.8\textwidth]{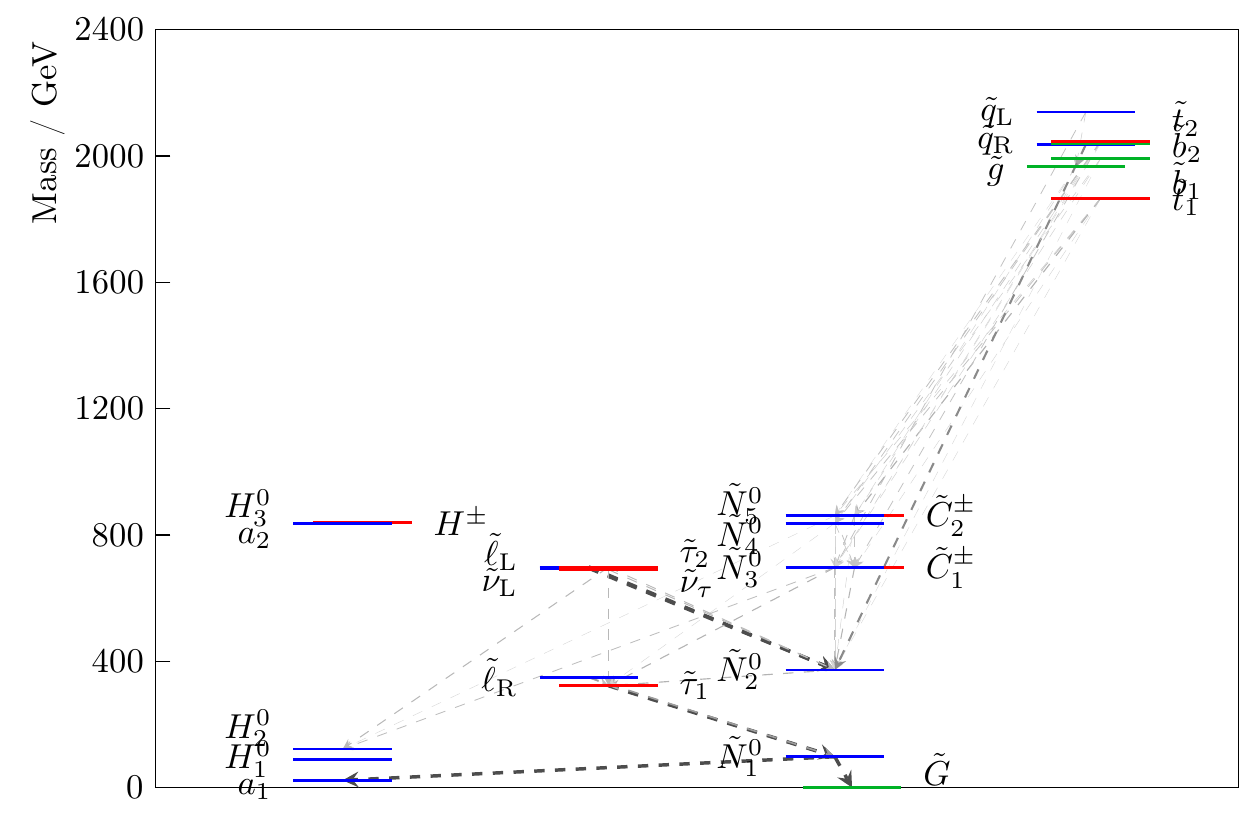}
\begin{tabular}{cccccccccccc}
 $m_{H_1^0}$ &$m_{H_2^0}$ &$m_{a_1}$ &$m_{\tilde{N}_1}$  &  $m_{\tilde{N}_2}$ &
$m_{\tilde{e}_1}$ & $m_{\tilde{\tau}_1}$ & $m_{\tilde{g}}$ & $m_{\tilde{u}_R}$ & $m_{\tilde{b}_1}$ &
 $m_{\tilde{t}_1}$ & $m_{\tilde G}$ \\
\hline
 90 & 122.1  & 23 & 97.6 & 373 & 348 & 323 & 1966 & 2045 & 1992 & 1866 & $5 \times
10^{-8}$ 
\end{tabular}
\end{center}

\caption{\label{fig:specP0} Spectrum and the more likely sparticle
  decays of benchmark point P0: $\xi=0.01$,
  $\lambda(M_{SUSY})=0.009$, $M=1.4 \times 10^6$ GeV, $\tilde m=863$
  GeV and $\tan \beta=28.8$.  Decays into sparticles which have a
  branching ratio greater than 10$\%$ are displayed by the arrows.
  The figure was produced with the help of {\tt
    PySLHA3.0.4}~\cite{Buckley:2013jua}. The table shows some more
  precise details of the spectrum of the benchmark point P0. All
  masses are listed in GeV units and the lightest neutralino has a
  decay length $c \tau_{\tilde N_1}=99$ mm.}
\end{figure*}

Here, we discuss the limits from the LHC, both in\linebreak prompt and displaced
searches.  Like minimal gauge mediation in the MSSM, the DGS
model also needs large radiative SUSY corrections to obtain the
correct Higgs mass (although the singlet-Higgs mixing helps). In order to be concrete, 
we choose to study a benchmark point P0, whose spectrum (as generated by
\textsc{NMSSMTools 4.9.2}~\cite{Ellwanger:2005dv,Ellwanger:2006rn}) is shown in
Figure~\ref{fig:specP0}. P0 has a SM-like Higgs in the vicinity of the
measured mass at 125 GeV (allowing for a 3 GeV theoretical uncertainty
in its prediction) and a lighter CP even Higgs at 90 GeV that couples
with a reduced strength (compared to a SM Higgs) to $Z$-bosons,
commensurate with a 2$\sigma$ LEP excess. In addition, the lightest
singlet-like pseudoscalar $a_1$ has a mass of 23 GeV
and the singlino-like NLSP $\tilde{N}_1$ has a mass of 98 GeV.

We generate event samples with \textsc{Pythia 8.2}~\cite{Sjostrand:2014zea}, using \textsc{FastJet 3.1.3}
\cite{Cacciari:2011ma} for jet reconstruction. The ATLAS models we wish to validate, described 
in section~\ref{DVValidation}, are generated with \textsc{SOFTSUSY 3.6.1}~\cite{Allanach:2001kg,Allanach:2013kza} to 
calculate the spectra and \textsc{SDECAY 1.5}~\cite{Muhlleitner:2003vg} to generate
the decays, communicating the spectrum and decay
information via SUSY Les Houches Accord (SLHA)
files~\cite{Skands:2003cj,Allanach:2008qq}. 

To take into account
the size of the detector, we consider a cylinder with radius $r=11$~m and
length $|z|=28$~m, corresponding to the ATLAS inner
detector~\cite{Aad:2009wy}. 
It is possible for a neutral particle that decays outside the inner detector
to form trackless jets. However, it is difficult to model the detector
response to these and so we consider them to be beyond the scope of this
study.  Any particle that decays outside the inner detector is therefore considered to
be stable for all intents and purposes.  The detector response for measurement of jet $p_T$ is modelled as follows\footnote{We find 
inconsistent results from standard detector simulation programs
leading us to believe that the presence of DVs
interferes with the standard reconstruction.}. 
The jet momentum is smeared by a gaussian with resolution of 20\% of energy for $E_\mathrm{jet} < 50$~GeV, falling linearly to 10\% up to 100 GeV and then a flat 10\%.  A further scale correction of 1\% is applied for jets with $|\eta| < 2$ and 3\% for those with higher $|\eta|$.

With this parameterisation, we are able reproduce the cut flows  for
the ATLAS 0-lepton + jets + missing transverse energy\footnote{We prefer to
  use the more accurate  
descriptor $\ptmiss = |\vecptmiss|$ than the `$\etmiss$' quoted by the ATLAS
analyses referred to in this study.} ($\ptmiss$) analyses and the efficiencies
are validated against published results for benchmarks provided in the
ATLAS analysis documentation. Further fiducial and material cuts required
for tracks in the DV studies are explained in
section~\ref{DVValidation}.

\section{Prompt SUSY searches \label{sec:prompt}}

In order to determine constraints on the gluino mass in our model, we
focus on the 0-lepton + 2-6 jets + $\ptmiss$ search~\cite{Aad:2014wea,Aaboud:2016zdn} which is the
most sensitive search for benchmark P0.  However, to investigate the response of our model to
dedicated SUSY searches, we deform it by moving on a line into the
phenomenological next-to-minimal model space (pNMSSM): for instance, we
vary the gluino mass soft parameter $M_3$ while keeping all other
weak-scale parameters fixed. The spectrum, decays and lifetimes are recomputed
at each point: to 
first order, only the gluino mass changes, but there are small loop-level
effects on other masses. Since this deformation breaks the gauge mediated
relation between 
the gaugino masses, we are deviating from the gauge mediated limit by doing
this. This is a simple choice where we can change only one parameter;
  we could have equally made a different choice where we vary several
  weak-scale parameters - trying to preserve some of the gauge mediated
  relations. Keeping within the NMGMSB model itself was not an option however,
since a highly non-trivial multi-dimensional manipulation of parameters was
required, which ended in some other phenomenological bound being violated.
Our approach 
is mainly phenomenologically motivated, essentially to study the gluino mass
bounds in the context of the very peculiar structure of singlino-like NLSP and
gravitino LSP (with squarks decoupled). Nevertheless one might imagine a
possible extension of the DGS scenario with additional sources for the Higgs
mass that allow to lower the overall scale of sparticle masses to the
investigated range.  

We have also sometimes, for the purposes of illustration only, changed 
the singlino decay length $c
\tau_{\tilde{N}_1}$ (while keeping all weak-scale parameters
fixed).  This deformation does not really constitute a consistent model, but
is used instead to understand some features that are present in consistent
models. When using this type of deformation we will refer to `tweaked'
parameters. 
We shall investigate the effect of varying the lifetime by scanning
over a lifetime range of  
$c\tau = [10^{-3},~10^4]$~mm for  $m_{\tilde g}\sim 1$~TeV.

\subsection{Current bounds from Run I and early Run II searches}
\label{sec:currentBounds}

\begin{table*}[t]
\begin{tabular*}{\textwidth}{@{\extracolsep{\fill}}lcccc@{}}

\hline
$\sqrt{s}$ & \multicolumn{2}{c}{8 TeV} & \multicolumn{2}{c}{13 TeV} \\
Signal Region & {\tt 4jt-8}  & {\tt 6jt-8} & {\tt 4jt-13} & {\tt 6jt-13} \\
\hline
$\ptmiss$/GeV $>$  & 160 & 160 & 200 & 200  \\
\hline
$p_T(j_1)$/GeV $>$ & 130 & 130 & 200 & 200 \\
$p_T(j_2)$/GeV $>$ & 60 & 60 & 100 & 100 \\
$p_T(j_3)$/GeV $>$ & 60 & 60 & 100 & 100 \\
$p_T(j_4)$/GeV $>$ & 60 & 60 & 100 & 100 \\
$p_T(j_5)$/GeV $>$ & - & 60 & - & 50 \\
$p_T(j_6)$/GeV $>$ & - & 60 & - & 50 \\
\hline
$\Delta \phi (\mathrm{jet}_{1,2,3}, \vecptmiss)_{min} > $ &  \multicolumn{4}{c}{0.4} \\ 
\hline
$\Delta \phi (\mathrm{jet}_{j>3}, \vecptmiss)_{min} > $ & \multicolumn{4}{c}{0.2} \\ 
\hline
$\ptmiss / m_\mathrm{eff}(N_j) > $ & \multicolumn{2}{c}{0.25} &
\multicolumn{2}{c}{0.2} \\ 
\hline
$m_\mathrm{eff}$(incl.)/GeV $ > $ & 2200 & 1500 & 2200 & 2000 \\
\hline
\hline
$\sigma^{\rm obs}_{95}$ (fb) &  0.15 & 0.32 & 2.7 & 1.6 \\
\hline
\end{tabular*}
\caption{\label{tab:cuts} The cuts for more sensitive signal regions from the
  0-lepton + jets + $\ptmiss$ searches at 8 TeV~\cite{Aad:2014wea} and 13
  TeV~\cite{Aaboud:2016zdn} runs and 95\% observed upper limits on a
  non-Standard Model contribution $\sigma^{\rm
    obs}_{95}$. The limit $\sigma^{\rm obs}_{95}$ has not been unfolded, and so
  should be applied to the production cross-section times branching
  ratio times acceptance. The jets $j$ are ordered in decreasing
  $p_T$. The effective mass, $m_\mathrm{eff}$(incl.) is defined to be the scalar sum of
  $p_T$'s of all jets with $p_T > 40 (50)\, {\rm GeV}$ for $\sqrt{s}=$8 (13) TeV plus the
  missing 
  transverse momentum $\ptmiss$, $m_\mathrm{eff}(N_j)$ is the scalar sum of
  $p_T$'s of $N_j$ hardest jets ($N_j=4$ for {\tt 4jt-X} and $N_j=6$ for {\tt
    6jt-X}) plus $\ptmiss$ and $\phi$ is the azimuthal angle
  around the beam.}
\end{table*}

In the NMGMSB model under study,
the squarks (including the third generation squarks) are usually
heavier than the gluino, resulting in three-body decays\linebreak through
off-shell 
squarks of the form
$\tilde g \rightarrow q \bar q \tilde N_1$, where $\tilde N_1$ is
mostly singlino-like, followed by the potentially displaced decay $\tilde N_1
\rightarrow \tilde G a_1 \rightarrow \tilde G b \bar b$. 
Although the last step
in the decay chain always ensures the 
presence of $b$'s in the final state, they are usually too soft to
satisfy the requirements of current $b-$jet + $\ptmiss$ searches.  
We find that $b-$jet searches only become sensitive when the mass of the gluino is high enough that decays into third generation squarks dominate and their decays into top/bottom quarks (see an example event topology in Figure~\ref{fig:feyn}) result in high-$p_T$ $b$-jets. Therefore these searches are never relevant for our benchmark P0 and the corresponding pNMSSM line that has even lower gluino masses.
Note however, that even when the gluino mass is high, the gluino branching fraction into $b$'s
is still only about 20\% and is often accompanied by vector bosons in
the final state. These sometimes produce leptons, which take events
out of the 0-lepton + multi-jets + $\ptmiss$ selection.  As a result of
the above considerations, the sensitivity is much lower than that from
simplified models producing hard $b$-jets and missing transverse
momentum.  We find that the sensitivity in the simplest
0-lepton + jets + $\ptmiss$ searches is greater than that of searches
involving $b$s even at high gluino masses. The signal regions (i.e.\ the
labelled sets of cuts) defined by 
ATLAS that have the highest sensitivity are the {\tt 4jt-8} and {\tt 6jt-8}
signal regions (relevant for 8 TeV collisions) and the {\tt 4jt-13}
and {\tt 6jt-13} signal regions (relevant for 13 TeV collisions).  We
reproduce the cuts in these signal regions in Table~\ref{tab:cuts}
along with the observed upper limits on production cross section at
the 95$\%$ confidence level (CL).

\begin{figure*}[t]
\begin{center}
\includegraphics[width=0.8\textwidth]{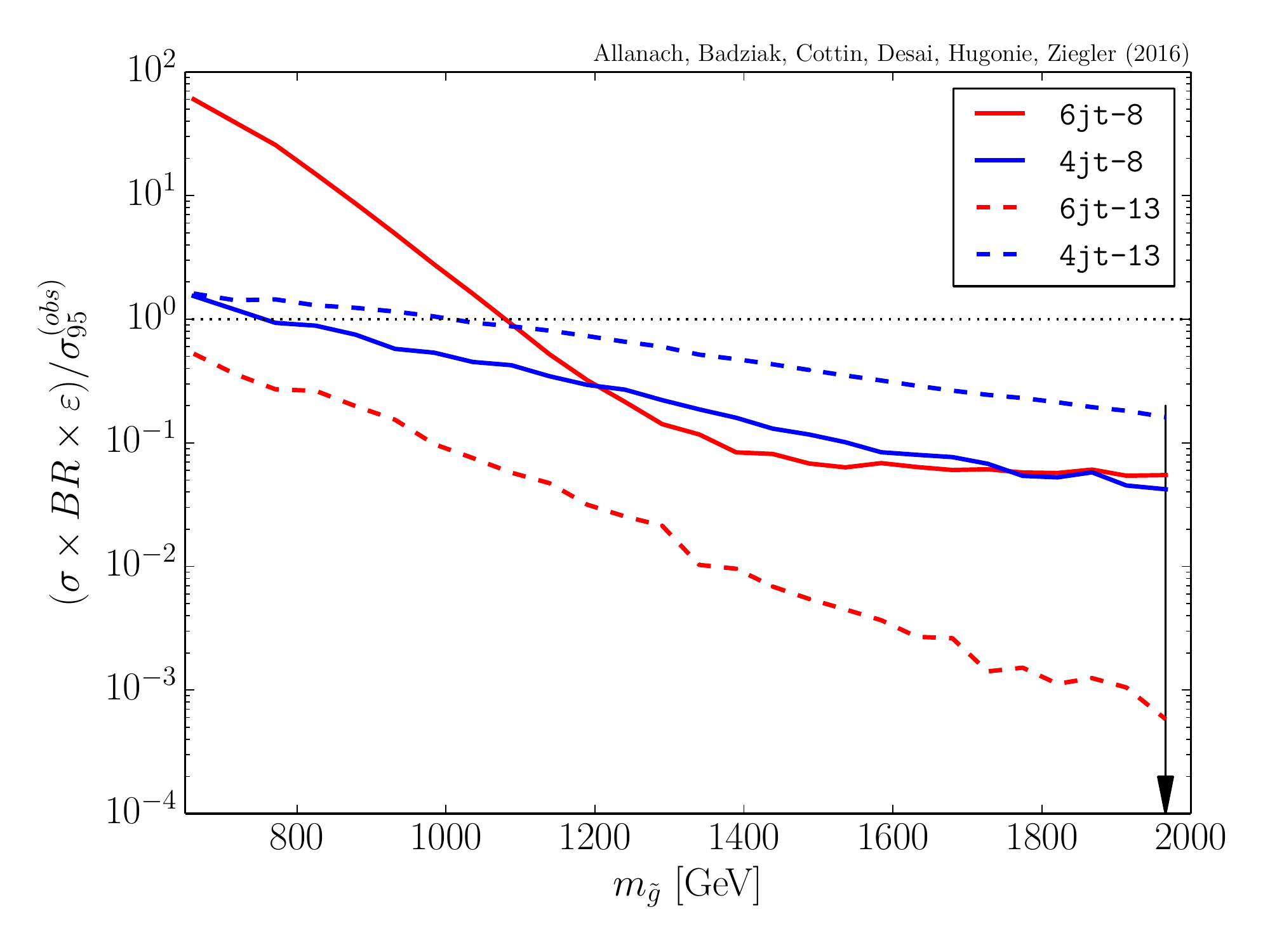}
\end{center}

\caption{95$\%$ lower limits on the gluino mass from Run I and
  Run II jets + $\ptmiss$ searches. We show the ratio of the predicted
  gluino cross-section times branching ratio times acceptance ($\sigma
  \times BR \times \varepsilon$) to the 95$\%$ upper bound on signal
  cross-sections determined by ATLAS, for a scan-line based on
  benchmark P0 ($\tilde N_1$ lifetime of $c \tau_{\tilde{N}_1} =
  99 \, {\rm mm}$). The horizontal dotted line shows the exclusion
  limit at $r=(\sigma \times BR \times
  \varepsilon)/\sigma_{95}^\mathrm{obs}=1$. The arrow shows the position of
  our benchmark P0 in NMGMSB, whereas elsewhere we are strictly in pNMSSM
  parameter space.
   \label{fig:gluinolimit}}
\end{figure*}

Since in the DGS model the gluino sets the overall mass scale of the sparticle spectrum, we 
therefore present our results in the form of bounds on the gluino mass.  We define 
the signal strength ratio $r_{95}$ as the ratio of the
predicted sparticle signal passing the selection cuts in a particular
signal region to the 95\% CL upper limit on the cross section in that
region. Thus $r_{95}=1$ is {\em just}\/ ruled out to 95$\%$ CL,
$r_{95}>1$ is ruled out whereas $r_{95}<1$ is allowed at the 95$\%$
CL. The signal region is always chosen to be the one giving the best
{\em expected}\/ exclusion. Figure~\ref{fig:gluinolimit} shows the signal 
strength ratio for varying gluino mass based on the pNMSSM line described in
section~\ref{sec:prompt}.  The most sensitive signal region for gluinos
from $0.9-1.2$ TeV is the {\tt 6jt-8} region (6 hard jets, tight cuts)
at $\sqrt{s}=8$ TeV. For higher gluino masses up to 2 TeV, the
sensitivities of the {\tt 6jt-8} and {\tt 4jt-8} signal regions are
similar.  The presence of a long-lived singlino which may decay
within the detector leads to another possible signature --- that of
DVs, which we shall explore in the next section. We see
from the figure that the bound from Run I at 95$\%$ is $m_{\tilde g}>$
\runIbound{}, where the {\tt 6jt-8} line intersects $r_{95}=1$.

\begin{figure*}[t]
\begin{center}
\includegraphics[width=0.8\textwidth]{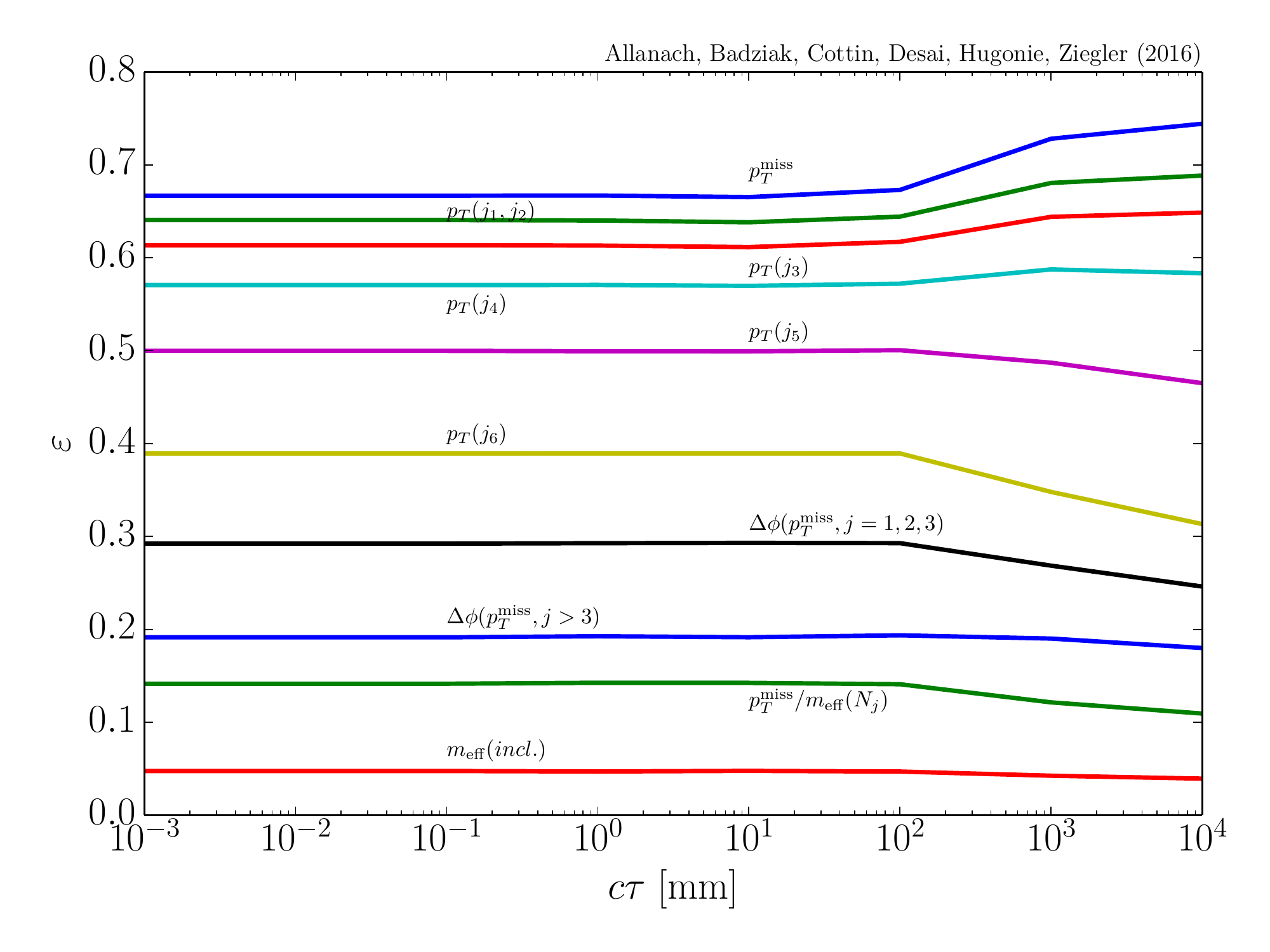}
\end{center}

 \caption{The dependence of overall efficiency on lifetime in the
  signal region {\tt 6jt-8}.  We find that the strong dependence on
  $\ptmiss$ is strongly anti-correlated with the cuts on jet
  $p_T$, resulting in a fairly small dependence of the efficiency on $c\tau$
  after all cuts.  The curves 
  correspond from top to bottom to the cuts in Table~\ref{tab:cuts} and the corresponding variables are shown in the plot.
  \label{fig:lifelimit}}
\end{figure*}

In 2015, ATLAS analysed 3.2~fb$^{-1}$ of integrated luminosity
collected at the higher centre of mass energy of 13 TeV, not
observing any significant signal for sparticle
production. We see that at 13 TeV, the {\tt 4jt-13} cuts are more
sensitive to our pNMSSM model than the {\tt 6jt-13} cuts for any
value of the gluino mass, mainly because the $p_T$ requirements
are not satisfied by the jets from singlino decay products (which give
$N_\mathrm{jets} > 4$). We see that the early Run II data from 2015
constrained $m_{\tilde g}>$ 1000 GeV, not as sensitive as the Run I
limit. Later, we shall examine the expected sensitivity from Run II
with 100~fb$^{-1}$ of integrated luminosity.

In Figure~\ref{fig:lifelimit}, we show how
changing the lifetime of the $\tilde N_1$ affects the cut acceptances
(shown here for the signal region {\tt 6jt-8}).  The lines should be
read in one-to-one correspondence from top to bottom with the cuts
listed in Table~\ref{tab:cuts} (except the second line which is the combined 
efficiency for $p_T(j_1)$ and $p_T(j_2)$).  Thus, the first line corresonds
to the cut $\ptmiss > 160$~GeV and the last to $m_\mathrm{eff} >
1500$~GeV.  As expected, when the singlino is stable, we find large
missing energy as all the momentum it carries is invisible.  As the
lifetime decreases, more and more singlinos decay within the detector
volume resulting in a flat efficiency below lifetimes of 10 mm.  This
gain in efficiency for long-lived $\tilde N_1$ is somewhat diluted
once we demand jets with high $p_T$.  In particular, once we demand
$N_\mathrm{jets} > 4$, the efficiency is lower for stable singlinos, since
the extra hard jets mainly come from decay products of the $\tilde
N_1$. However, this downturn is balanced by the requirements on
$\Delta \phi$ between the jets and the missing momentum
$\ptmiss$, since the presence of more jets in the final
state makes it harder to satisfy this cut. Finally, after all cuts, we
find that the efficiency is rather flat across all lifetimes.
Therefore the gluino mass limits presented above may be considered
fairly robust for the model studied here.

\subsection{Future search reach of prompt searches \label{sec:promptfututre}}

\begin{figure*}[ht]
\begin{center}
\includegraphics[width=0.8\textwidth]{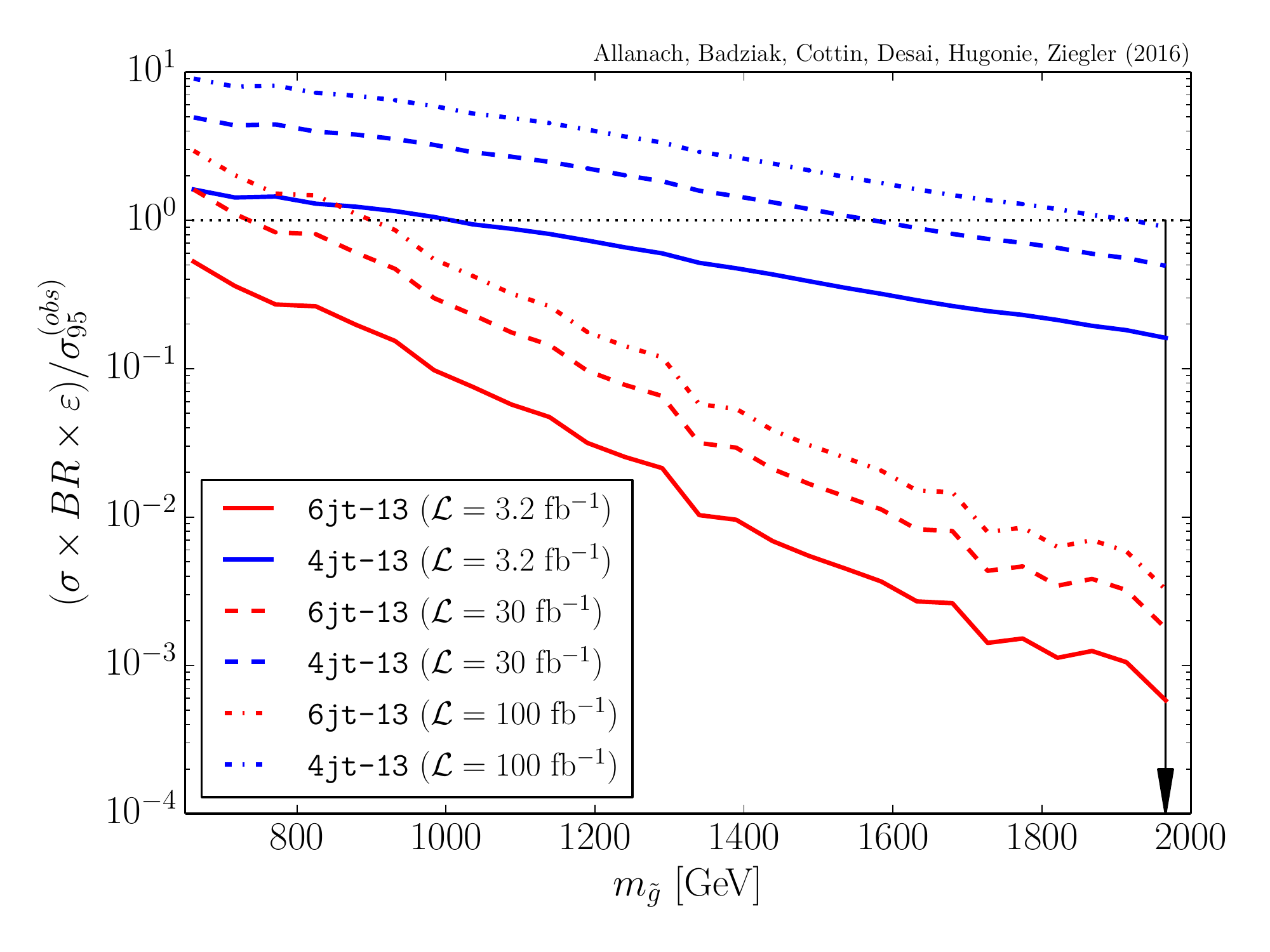}
\end{center}

\caption{95$\%$ CL limits the gluino mass from the $\sqrt s =13$ TeV
  jets + $\ptmiss$ searches and projected sensitivity with higher
  luminosity in the pNMSSM model. The signal regions {\tt 4jt-13} and
  {\tt 6j-13} are defined in Table~\protect\ref{tab:cuts}. The arrow shows the position of
  our benchmark P0 in NMGMSB, whereas elsewhere we are strictly in pNMSSM
  parameter space.
\label{fig:gluinolimit13}} 
\end{figure*}

We now estimate what the future might bring for discovery or exclusion
of the pNMSSM model from the LHC.  In Figure~\ref{fig:gluinolimit13},
we re-display the current limits on the gluino in pNMSSM from the 13
TeV run, as the gluino mass is changed for the {\tt 6jt-13} and {\tt
  4jt-13} signal regions. The solid lines show the current lower limit
from the 2015 run of 1000 GeV. For our model, the {\tt 4jt-13} region
performs better than the {\tt 6jt-13} region. Model sensitivity (ignoring
systematic errors) is equal to the number of signals events $S$
divided by the square root of the number of background events
$B$. Since $S \propto$ the total integrated luminosity ${\mathcal L}$,
and $B \propto {\mathcal L}$, the sensitivity scales $\propto
\sqrt{\mathcal L}$. Thus, we expect $\sigma_{95}^\mathrm{obs} \propto
1/\sqrt{\mathcal L}$. Using this dependence, we scale the ${\mathcal
  L}=3.2$~fb$^{-1}$ lines to 30~fb$^{-1}$ and 100~fb$^{-1}$ to show
the projected sensitivities in the figure. We see that with
100~fb$^{-1}$ and 13 TeV centre of mass collision energy, the LHC can
reach up to 1900 GeV gluinos.

\section{Searches with displaced vertices \label{sec:displaced}}

DV searches are especially challenging due to the complication 
of taking into account time of flight and assigning tracks originating far 
away from the primary interaction point to the correct event.
Reconstruction of such decays therefore becomes more difficult beyond the
pixel layers. Nevertheless, these searches have an extremely low
background as there are no irreducible contributions from the SM.
Some recent reinterpretation of LHC displaced searches
can be found in
Refs.~\cite{Liu:2015bma,Csaki:2015uza,Cui:2014twa,Csaki:2015fba,Evans:2016zau,Coccaro:2016lnz,delaPuente:2015vja}.
Displaced signatures have received far less attention in the
literature as compared to prompt signatures because they are difficult
to model, and because they tend to be rather model specific. Furthermore, modelling
the detector's response to DVs is a difficult task, as
we shall illustrate. Validation is therefore essential in order to
tell how good or bad a job of modelling the response we achieve.
Refs.~\cite{Liu:2015bma,Cui:2014twa} used truth information to
identify displaced decays. Our work goes further by fully detailing
the steps of reconstruction for DVs, in a similar way
to Ref.~\cite{Csaki:2015uza}, but here we determine an explicit
functional form for the tracking efficiency, which is needed to be
able to model the efficiencies from the experiments to a reasonable
(if somewhat rough) level. 

\subsection{Validation of Run I displaced vertex searches \label{DVValidation}}

 In the absence of publicly available multi-dimensional, model-independent efficiency maps 
 for the reconstruction efficiency of a DV, 
 we make use of the efficiencies published for specific models and construct a function 
 that approximately simultaneously reproduces them.
 The ATLAS DV + jets search~\cite{Aad:2015rba} has been interpreted in the context of two
 General Gauge Mediation (GGM) and several $R-$parity violating
 supersymmetry (RPV) simplified models. Of these, the ones most relevant to signatures predicted 
 by the DGS model (where we expect only jets from the  DV) are the two GGM model 
 benchmarks and one RPV benchmark where a displaced
 neutralino decays through a non-zero $\lambda'_{211}$ to light quarks
 and a muon.

\begin{table*} 
\begin{center}
\begin{tabular*}{0.8\textwidth}{@{\extracolsep{\fill}}ll@{}}
\hline
DV jets     & 4 or 5 or 6 jets with  $|\eta| <2.8$ and $p_{T} > 90, 65,
55$ GeV, each.\\ 
\hline
DV reconstruction   & DV made from tracks with $p_{T}>1$ GeV, $|\eta|<2.5$ and $|d_{0}|>2$ mm, satisfying a\\ 
                    & tracking efficiency given by equation \ref{eq:trackEff}. Vertices within 1 mm are merged. \\
\hline 
DV fiducial         & DV within $4$ mm $<r_{DV}<300$ mm and $|z_{DV}|<300$ mm.
\\  
\hline 
DV material         & No DV in regions near beampipe or within pixel layers: \\ 
                    & Discard tracks with $r_{DV}/\mathrm{mm} \in \{[25,38],
                    [45,60], [85,95], [120,130]\}$.\\
\hline 
$N_{\rm trk}$       & DV track multiplicity $\geq 5$. \\
\hline 
$m_{DV} $           & DV mass $>10$ GeV. \\
\hline 
\end{tabular*}
\end{center}
 
\caption{\label{tab:cutflow_ATLAS} Our implementation of cuts applied in the ATLAS
  multi-track DV + jets search, from Ref.~\cite{Aad:2015rba}.}
\end{table*}
The ATLAS DV + jets cuts are summarised in Table~\ref{tab:cutflow_ATLAS}.
The ATLAS analysis re-runs the experiment's
standard tracking algorithms on events passing the trigger in order to determine the efficiency for
the displaced tracks. Given the fact that we do not have access to
such algorithms, we assign each track a reconstruction probability depending
on its $p_T$  and the \emph{true} co-ordinates of its displaced origin.  The
functional form found to reproduce the efficiencies for the three
benchmark models is given by 
\begin{eqnarray}
\varepsilon_\mathrm{trk} & = & 0.5\times(1-\exp(-p_{T}/[4.0\mathrm{~GeV}])) \nonumber \\
& & \times\exp(-z/[270\mathrm{~mm}]) \nonumber\\
& & \times \mathrm{max}(-0.0022\times{r_{\bot}/\mathrm{[1\mathrm{~mm}]}}+0.8,0),
\label{eq:trackEff}
\end{eqnarray}
where $r_{\bot}$ and $z$ are the transverse and longitudinal distance
of the track's production vertex (for details of their definition, see
Appendix~\ref{sec:def}). 

\begin{figure*}[ht]
\begin{center}
 \includegraphics[width=\textwidth]{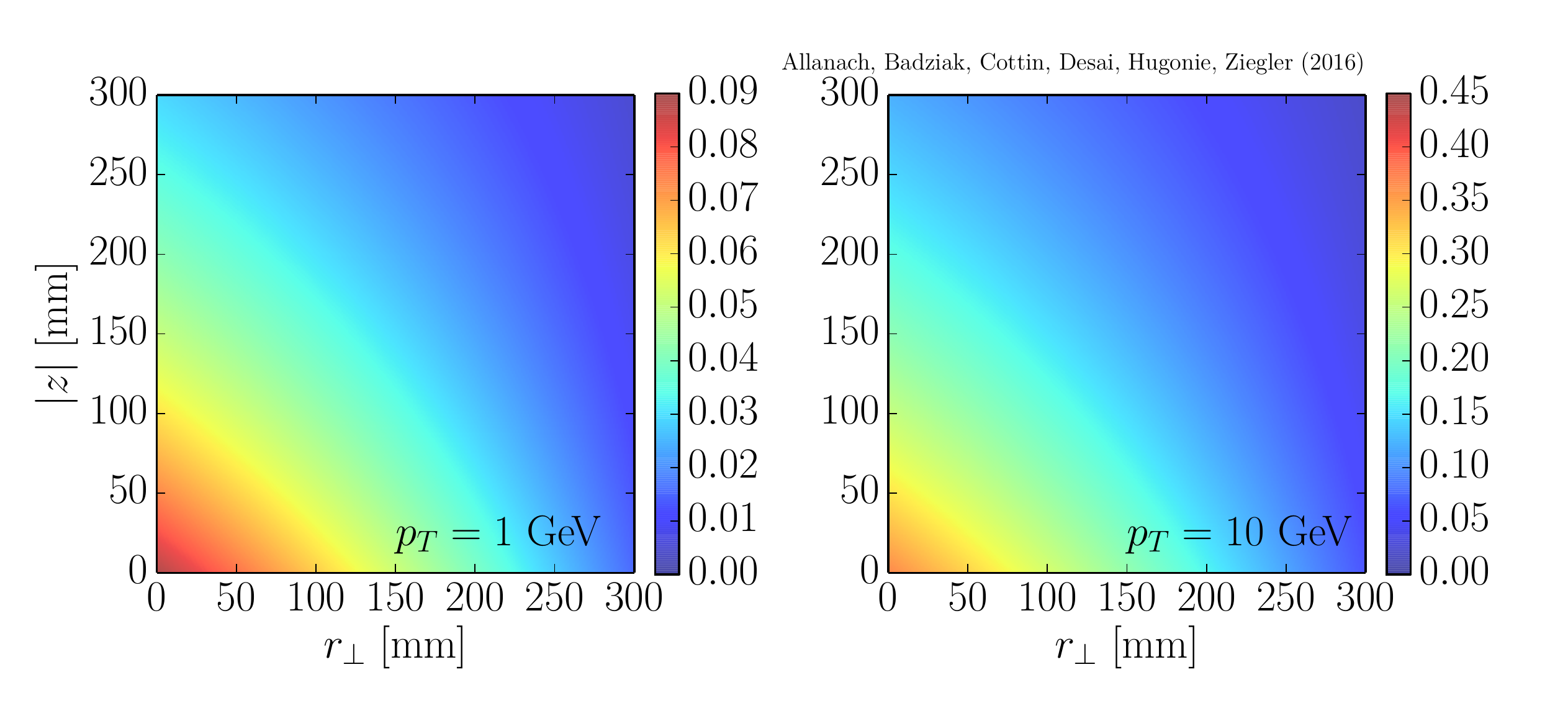}
\end{center}

\caption{\label{fig:contEff} Contours of track efficiency as a function of
  $r_\bot$ and $|z|$, for a $p_{T}$ value fixed at 1 GeV (left) and 10 GeV
  (right). Note the difference scales of track efficiency (labelled by the
  legend at the right hand side) of each panel.}
\end{figure*}

We pick this particular parameterisation of
the tracking efficiency after trying several functional forms and
varying the constants, picking the one that had the best goodness of
fit statistic ($\chi^2$) for the three models combined that we
validate against (at various different values of lifetimes of the
decaying sparticle). Eq.~(\ref{eq:trackEff}) is not expected to be
perfect by any stretch: it is a simple, universal and factorised
form for the track efficiency that is a rough approximation. The
overall $\chi^2$ statistic did not indicate a particularly good fit,
however inspection by eye showed that the shapes of the efficiency
curves were reasonable. We display contours of the function in
Figure~\ref{fig:contEff}.

The efficiency for reconstructing a multi-track DV is
highly dependent on track reconstruction and track selection, as
detailed in Ref.~\cite{Aad:2015rba}. These are affected by several
factors:
\begin{itemize}
\item{The impact parameter $d_{0}$ of the track: the efficiency for
  reconstructing tracks decreases with increasing values of $d_{0}$,
  since the density of fine instrumentation decreases.} 
\item{The mass of the long-lived particle: as the number of tracks
  originating from the vertex increases with increasing mass. At
  higher masses, missing some tracks may therefore still lead to the
  identification of a DV.}
\item{The energy of the long-lived particle: the higher the boost,
  the more tracks will have a small angle with respect to the
  flight direction of the long-lived particle and may therefore fail
  the minimal $d_{0}$ cut.}
\end{itemize}
We can therefore see that the vertex reconstruction efficiency will be
a non-trivial combination of different aspects of the track
reconstruction efficiency. For instance the vertex reconstruction
efficiency is worst at large radii, which is due to tracking
efficiency decreasing. Here, we focus only on assigning each track a
reconstruction probability dependent on some relevant variables, such as
the transverse and longitudinal distance of the production vertex of
the track and its transverse momentum. The particular choice of the
variables in equation \ref{eq:trackEff} (for instance the fact we use
$r_\bot$ instead of $d_{0}$) was made because we found these fitted the
best across the three different signal models out of a few different
simple factorised functional forms that we tried.

\begin{figure*}[htp]
\centering
\subfigure[GGM1]{\includegraphics[width=\columnwidth]{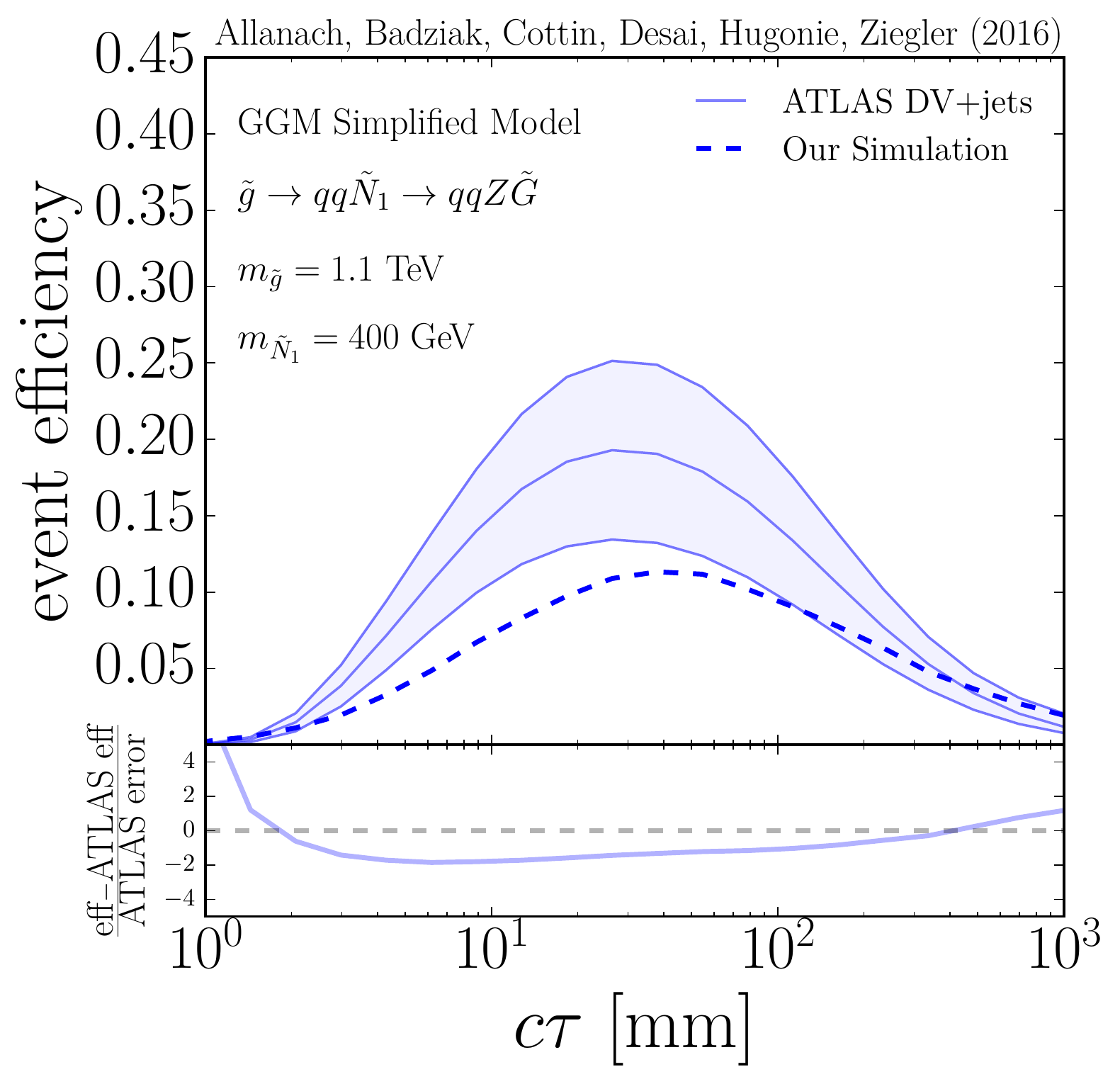}}
\subfigure[GGM2]{\includegraphics[width=\columnwidth]{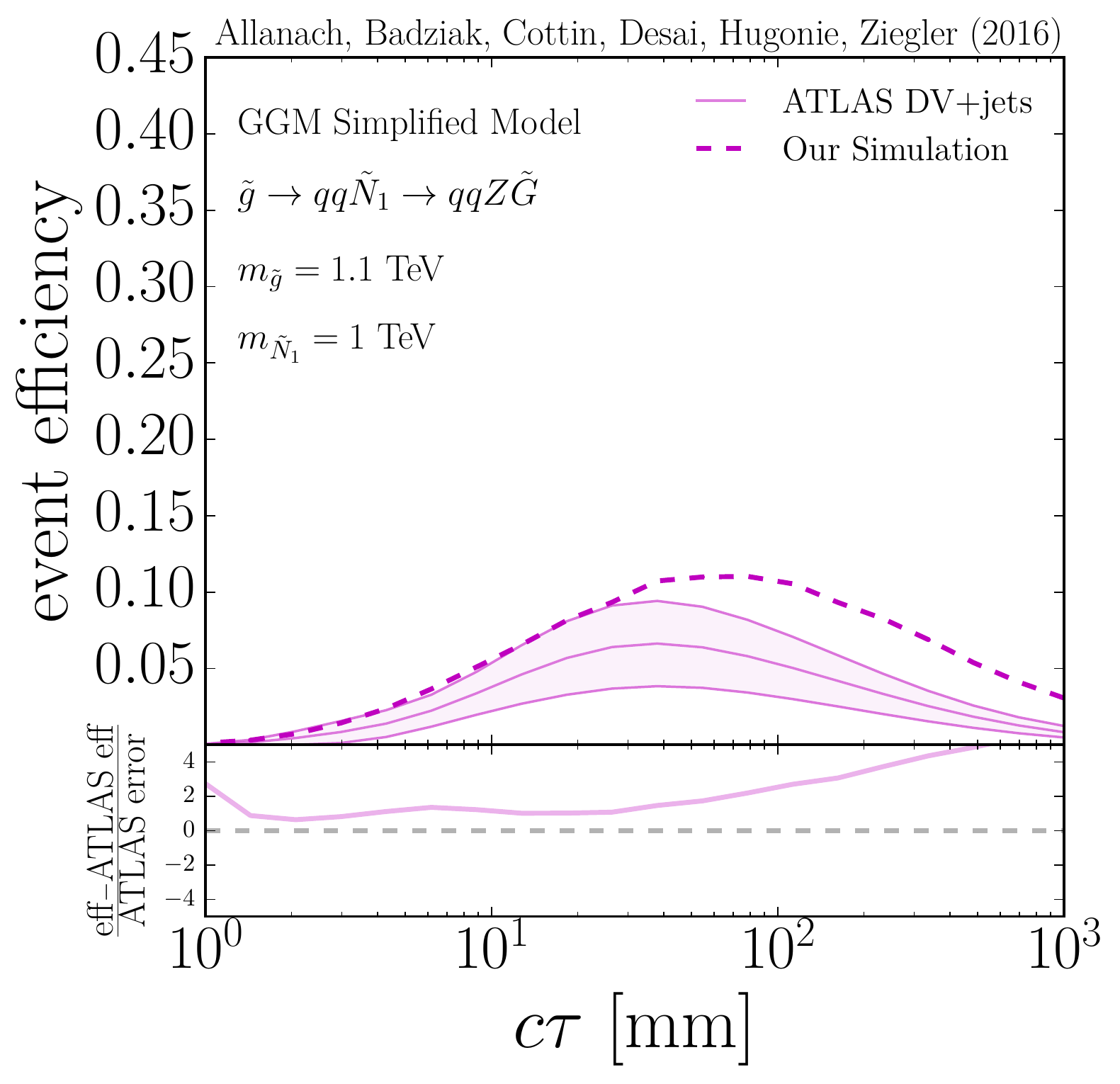}}
\subfigure[RPV]{\includegraphics[width=\columnwidth]{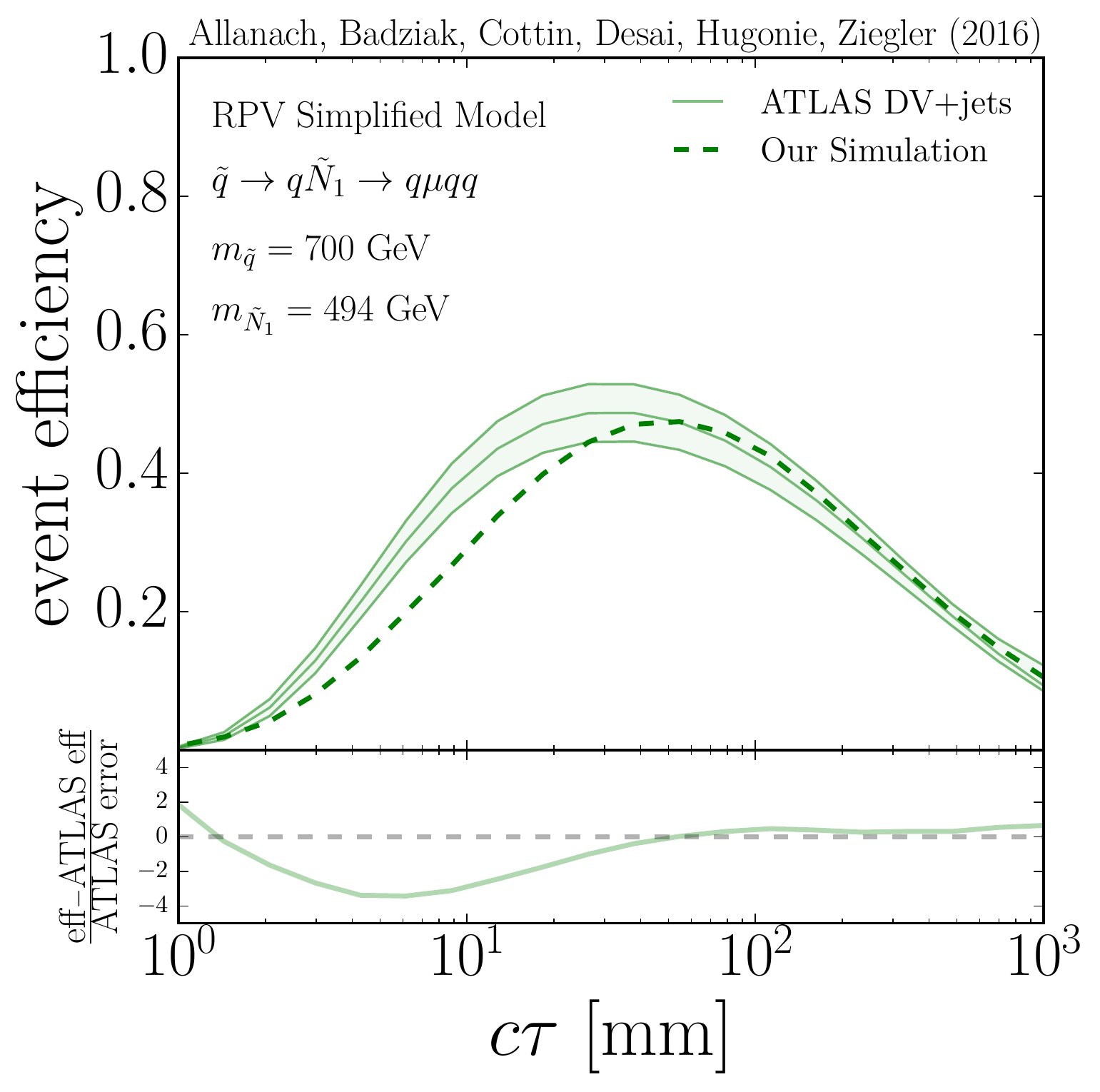}}
\caption{Validation of our DV + jets estimate of efficiency for three
  models against ATLAS's determination: (a) a simplified GGM model
  where a $1.1$ TeV gluino decays to a $400$ GeV neutralino, which in
  turn decays to a $Z$ and a gravitino, (b) a simplified GGM model
  where a $1.1$ TeV gluino decays to a $1$ TeV neutralino, which in
  turn decays to a $Z$ and a gravitino, (c) a simplified RPV model
  with a 700 GeV squarks decaying to a $500$ GeV neutralino, which
  subsequently decays through a non-zero $\lambda'_{211}$ coupling
  into a muon and two quarks.  Events are generated with $\sqrt{s} = 8
  $ TeV.  The bottom rectangle in each case shows the discrepancy
  between our estimate and ATLAS's, measured in units of the ATLAS
  error.
\label{fig:effVsCtau}}
\end{figure*}

We find that validating against only one of the benchmark models 
at a time leads to different best-fit parameters for each. Using three benchmarks 
for validation therefore gives us more confidence in applying the efficiency to our own model.
We believe this is a key improvement in our work. 
Choosing a functional form for the tracking efficiency in order to
recast displaced results has already been attempted in the
literature~\cite{Csaki:2015uza}.  Here we show the explicit functional form used, since 
knowing it is necessary to be able to reproduce our results. 
 Figure~\ref{fig:effVsCtau}
shows the validation of our simulation (dashed lines) for three
different ATLAS benchmarks, against the ATLAS determination (solid
lines). We see that the efficiency, while far from being perfectly
modelled by our function, is adequately modelled (within a few sigma)
for most of the range of lifetimes considered.
We could improve the above fits by including an additional selection
efficiency at the vertex level, as discussed above. We could also take
into account the topology of the different signatures. Ideally, a
parameterisation of the tracking efficiency should be validated against
all the $\sim 20$ signal benchmarks used in the ATLAS search, which is
beyond the scope of this paper and which we leave for a future work.

Undiscarded displaced tracks are input into our vertex reconstruction
algorithm, which compares and clusters the tracks'
origins\footnote{ATLAS performs a complicated vertex $\chi^2$ fit in
  order to {reconstruct DVs}. Here, we simply use the
  truth information to define the track's origin to be the point at
  which the $\tilde N_1$ decays, and start comparing the distance
  between tracks' origins to cluster them into vertices.}. If the
origins of two displaced tracks are less than 1 mm
apart, then they are clustered together into one DV. Picking 
the first track, we compute the $d$ value (i.e.\, the physical distance in the
laboratory frame $\sqrt{\Delta x^2 + \Delta y^2 + \Delta z^2}$) to each of the
other tracks, clustering tracks that have a small enough $d$ value to
the first track. Then we repeat for the next unclustered track and so on,
until each track is assigned to a single vertex. The ATLAS analysis
(and ours) combines vertices into a DV if they are less than $d=1$ mm
apart. The 
DV position is defined as the average position of all the track
origins in the cluster.

To ensure consistency of the vertex position and the direction of the
tracks, we require at least two tracks in the vertex to have
$\vec{d}\cdot\vec{p}>-20 $ mm, where we define $\vec{d}$ to be the
vector from the interaction point to the DV and $\vec{p}$ to be the
momentum of the displaced track. In the ATLAS analysis, DVs are vetoed
if they are reconstructed in high density material regions, since this
is the main source of background vertices.  We simulate this by requiring
$4$ mm $< r < 300$ mm and $|z|<300$ mm. We also require decay
positions of the DVs to not be inside any of the three ATLAS pixel
layers (our approximation to this DV material cut is shown in
Table~\ref{tab:cutflow_ATLAS})\footnote{Note that the material veto
  performed in the ATLAS analysis is far more complex than this, since
  ATLAS makes use of a 3D material map of the detector that we do not have
  access to.}. As the table shows, events are further selected if they
have at least one reconstructed DV with 5 tracks or more and a DV
invariant mass (computed assuming all tracks have the pion mass) of at
least 10 GeV.

\subsection{Run I sensitivity of displaced vertex searches \label{sec:DVruni}}

We now apply the simulation described in the previous
section to the DGS model benchmark P0. We find that the sensitivity of the
ATLAS study to our benchmark is extremely limited.  From the 8 TeV columns of
Table~\ref{tab:cutflow_DGSATLASDefault}, it is clear that the primary cause
for this is failure to satisfy the requirements $N_{\rm trk} \geq 5$ and  the
vertex mass cut $m_{\rm DV} > 10$~GeV. 
\begin{table*}
\begin{tabular*}{\textwidth}{@{\extracolsep{\fill}}lrrrr@{}}

\multicolumn{4}{c}{} \\ 
\hline
                    &$\sqrt{s} = 8 $ TeV  &                           &  $\sqrt{s} =13 $  TeV  &               \\
                    & $N$          & $\epsilon$ [\%]   &   $N$           & $\epsilon$ [\%]   \\ 
\hline  
All events          &    100000          &   100.             & 100000        &  100.               \\ 
DV jets             &     96963          &   97.              & 98306        &  98.3               \\ 
DV reconstruction   &     16542          &   17.1             & 16542        &  16.8               \\ 
DV fiducial         &     16459          &   99.5             & 16460        &  99.5               \\ 
DV material         &     16146          &   98.1             & 16210        &  98.5               \\ 
$N_{\rm trk}$       &       584          &   3.6              &   544        &  3.4                \\ 
$m_{\rm DV} $       &         4          &   0.7              &    3         &  0.6                \\ 
\hline 
%
\end{tabular*} 
\caption{\label{tab:cutflow_DGSATLASDefault}Numbers of simulated
  events $N$ and relative efficiencies $\epsilon$ (i.e.\ defined with
  respect to the previous cut) for our NMGMSB model (P0 benchmark) with
  $c\tau_{\tilde{N}_{1}}=99$ mm at $\sqrt{s}=8$ TeV and $\sqrt{s}=13$
  TeV for the ATLAS selection of cuts in Table~\ref{tab:cutflow_ATLAS}. }
\end{table*} 
This is due to the fact that the displaced jets are mainly $b-$jets.  The $b-$hadrons are themselves long-lived, and the neutral $B^0$ leaves no tracks before its decay.  The topology of this final state then has two further DVs, each with less than 5 tracks.  The ATLAS analysis does merge vertices 
(defined as having at least two tracks) that are within 1 mm of each other to possibly obtain a better vertex.  However, the $b-$hadrons are sufficiently long-lived so that the resultant vertices are almost always more than 1 mm apart\footnote{The ATLAS analysis~\cite{Aad:2015rba} also reports that the sensitivity is severely reduced if they use the RPV benchmark with $b-$quarks in the final state. An earlier work on displaced Higgs decays~\cite{Csaki:2015fba}, also shows how displaced $b$-quarks can be problematic, particularly given the $d<1$ mm requirement for merging vertices.}.  For the benchmark P0 for instance, the average displaced track efficiency is 0.06, and the average number of tracks coming from a displaced $b$ is 18.1 (after hadronisation, but before cuts).  Thus, on average, there are only 18.1$\times$0.06$=${1.2} visible tracks per displaced $b$.

\begin{figure*}[ht]
\centering
\includegraphics[width=\textwidth]{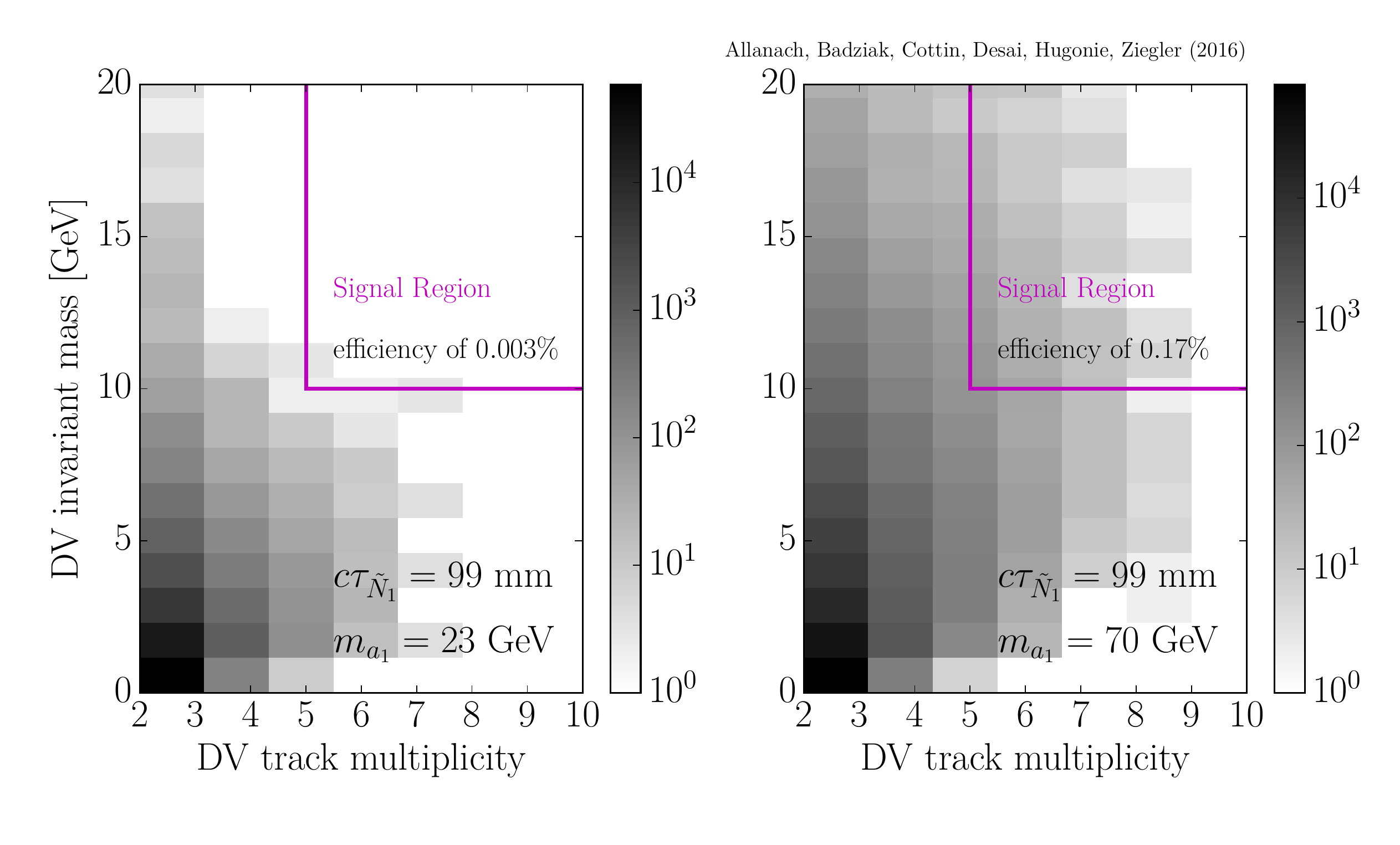}
\caption{DV invariant mass against number of tracks for
  the DGS benchmark P0 with $m_{\tilde{g}}=1.96$ TeV,
  $m_{\tilde{N}_{1}}=98$ GeV and $c\tau_{\tilde{N}_{1}}=99$
  mm. Gluinos and squarks only are generated with $\sqrt{s}=8$
  TeV. Events in the plot pass all of the DV cuts except for the last
  two, which define the boxed ATLAS signal region. The left-hand frame
  shows a scenario where $m_{a_{1}}=23$ GeV (in the DGS good-fit
  region, as in P0) and the right-hand frame has a tweaked
  $m_{a_{1}}=70$ GeV in the SLHA file (i.e.\ inconsistent with the soft
  parameters, which are left constant - for the purposes of illustration only).}
\label{mDVnTrk}
\end{figure*}

\begin{figure*}[ht]
\centering
\includegraphics[width=\textwidth]{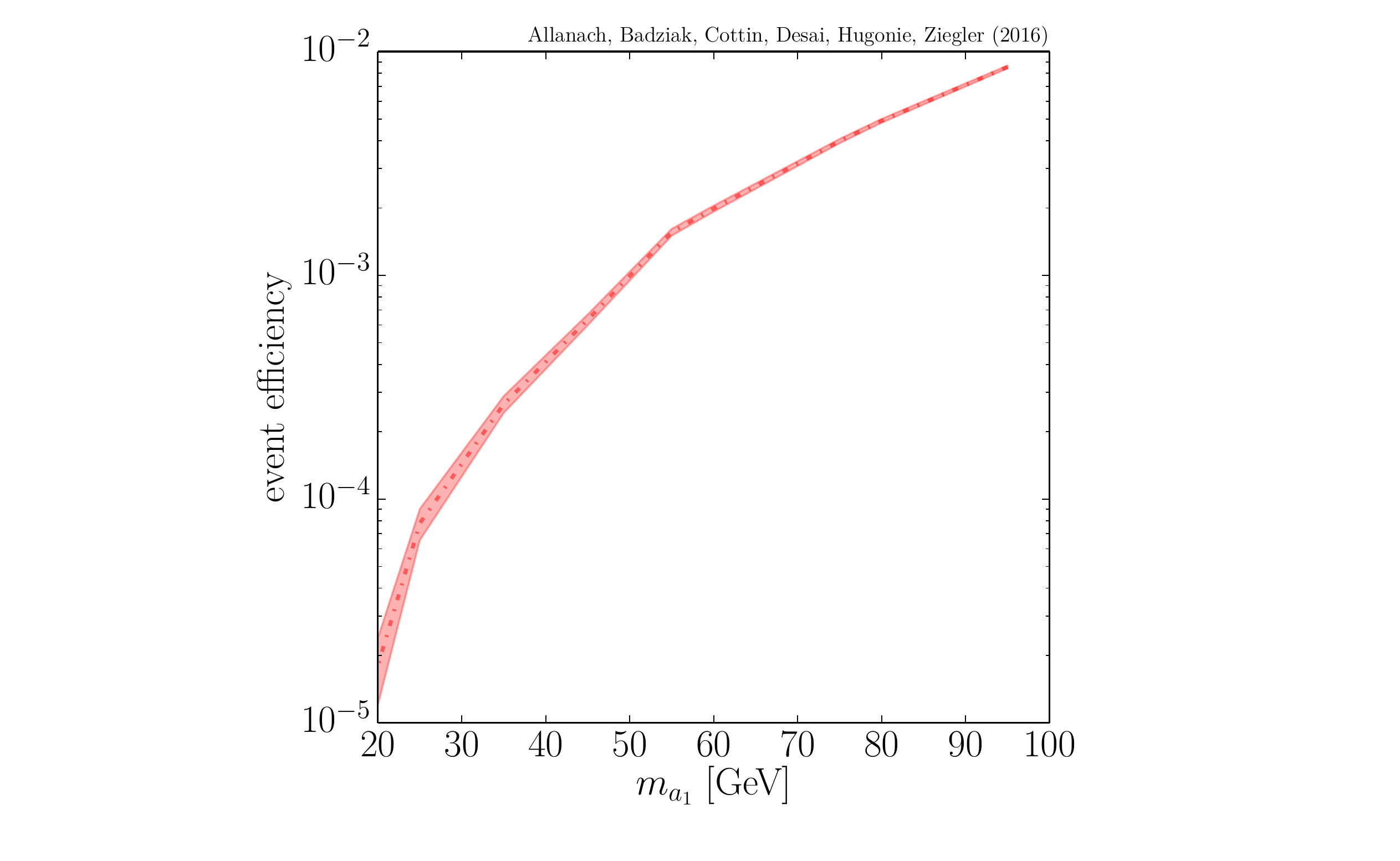}
\caption{Event efficiency against pseudoscalar mass for a DGS
  benchmark with $c\tau_{\tilde{N}_{1}}=99$ mm (our P0 benchmark). Events are generated
  with $\sqrt{s}=8$ TeV considering strong production. We have tweaked
  $m_{a_1}$ ``by hand'' in the SLHA files without changing soft parameters for
  the purposes of illustration.}
\label{effVsA1Mass}
\end{figure*}

A further consideration is the small mass of the $a_1$ which decays to $b \bar
b$ (23 GeV for the benchmark P0) since softer $b-$quarks means less radiation,
implying fewer tracks.  The distribution of track multiplicity versus
invariant mass is shown in Figure~\ref{mDVnTrk}.  One can see clearly from the
right panel that increasing the $a_1$ mass to 70 GeV (done \emph{ad hoc} for
the purposes of illustration) improves the sensitivity of the cuts by two
orders of magnitude.  A higher mass also means the resultant products are more
collimated and hence the $b$-hadron vertices are likely to be closer to each
other.  The improvement in efficiency with increasing $a_1$ mass can be seen
in Figure~\ref{effVsA1Mass}.

\subsection{Improving the sensitivity of displaced vertex searches \label{sec:DVimprov}}

Given the very low sensitivity of the DV searches, we shall now attempt 
to improve it by loosening the most restrictive cuts. 
Firstly, to catch DVs coming from two
$b-$quarks  
from the same $a_1$, we relax the requirement of maximum merging distance 
from 1 mm to 5 mm.  Further, we can also relax the last two cuts: track multiplicity and
invariant mass of the DV.

The background to the DV multi-track search comes from three sources --- heavy flavour
quark decays, interactions with material in the detector and the
accidental crossing of tracks, all of which have a low multiplicity of tracks and a small invariant mass of the
DV.  Thus, if we loosen these cuts to achieve better signal
efficiency, we also raise the background rate thus reducing
the signal to background ratio. However, given that our model has good sensitivity in the prompt $\ptmiss$-based channels,
background rates can be controlled by taking advantage of the
hard prompt signals that come in association with the DVs. Requiring a large
$m_\mathrm{eff}$ in the event 
would reduce backgrounds significantly. It may also be possible to increase the
sensitivity by loosening the DV cuts but requiring displaced jets to
have a muon inside them~\cite{Abazov:2009ik,Aad:2013txa}
(which often come from a $b$).  However, we do not consider this route here.

We now investigate the effect of applying prompt cuts used
in standard jets + $\ptmiss$ sparticle searches
on top of relaxed DV cuts.  This, of course, will have a lower signal
efficiency than purely applying the standard
jets + $\ptmiss$ cuts, which are already designed to remove
the SM background very effectively. Ideally, one would optimise the jets + $\ptmiss$ cuts along with the DV 
cuts to reach an overall best sensitivity. However, we have clear estimates of the background to the prompt channels from the analysis which serves as an upper bound to any DV contributions we may have from heavy flavour.  
Of course, the contributions from systematic sources cannot be bounded in this way, however, we can reasonably assume that the number of DVs from systematic sources is not biased by the hard cuts we place.

\begin{figure*}[ht]
\centering
\includegraphics[width=\textwidth]{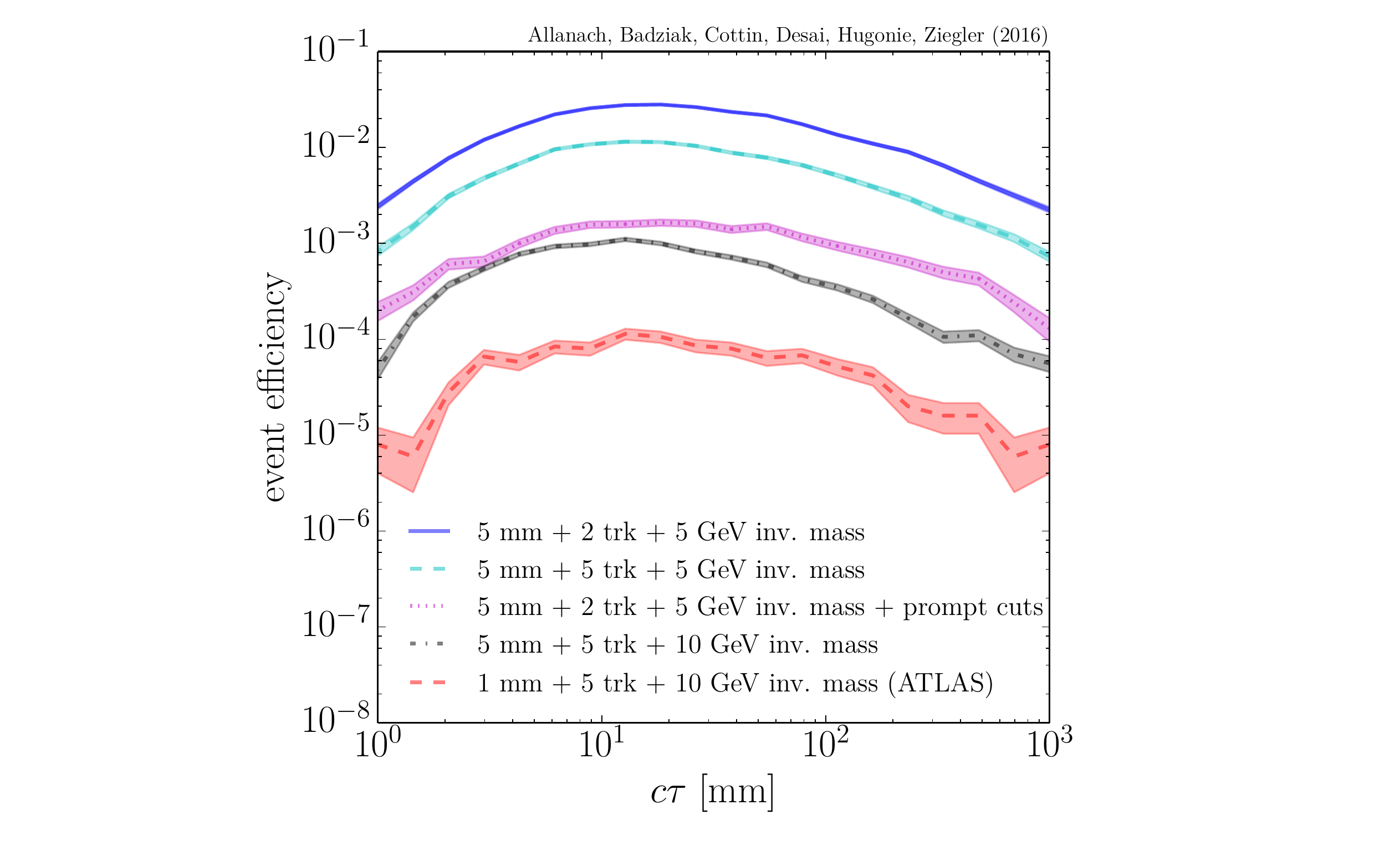}
\caption{Signal efficiencies of different sets of DV + jets analyses on
  a DGS benchmark with $m_{\tilde{g}}=1.96$ TeV,
  $m_{\tilde{N}_{1}}=98$ GeV, $m_{a_{1}}=23$ GeV (our P0 benchamark) against the lifetime
  of the long-lived singlino $c\tau_{\tilde{N}_{1}}$ (changed ``by
  hand" in the SLHA files without changing the soft terms for the purposes of
  illustration). Events are generated with $\sqrt{s}=8$ TeV,  
  considering gluino/squark production only. The bottom curve corresponds to
  the efficiency after the default ATLAS DV cuts. The top curve
  corresponds to the loosest of our selections in DV
  merging distance, track multiplicity and invariant mass. We also
  show different sets of cuts in between, including the inclusion of
  standard prompt cuts, as defined in the text.}
\label{effVsCtau_DGS}
\end{figure*}

At 8 TeV, we choose the ATLAS {\tt 6jt-8} signal region cuts described in
Table~\protect\ref{tab:cuts}, because they were found to have 
the highest sensitivity to our signal, as shown above.
Figure~\ref{effVsCtau_DGS} shows 
efficiency curves against lifetime for the NMGMSB model with the
default ATLAS DV analysis cuts and some choices of relaxed cuts.  This
includes (i) allowing $N_\mathrm{trk}$ to be $ \geq 2$ rather than $\geq 5$, (ii) increasing the vertex
merging distance from 1 mm to 5 mm, and (iii) lowering the vertex mass cut
from 10 GeV to 5 GeV. For comparison, we also show the response for the
original tight ATLAS DV ({\tt DVT}) cuts as well as our loose cuts ({\tt DVL})
for the {\tt 6jt-8} signal region.  
With this combination, we already achieve an improvement in
signal efficiency by a factor of ten. Without the {\tt 6jt-8} cuts, the
improvement is a factor of several hundred.  An optimised analysis will be
between these two limiting cases and may therefore be reasonably expected to
offer an improvement of two orders of magnitude or so. 

\subsection{Recommendations for displaced vertex searches at 13 TeV \label{sec:DVrunii}}
\begin{table*}
\begin{tabular*}{\textwidth}{@{\extracolsep{\fill}}lrrrr@{}}

\multicolumn{4}{c}{} \\ 
\hline
                     &$\sqrt{s} = 8 $ TeV  &                           &  $\sqrt{s} =13 $  TeV  &                        \\
                     & $N$          & $\epsilon$ [\%]   &   $N$           & $\epsilon$ [\%]   \\ 
\hline  
All events                                                        & 100000 & 100.  &  100000 &  100.   \\         
Prompt ${\ptmiss}^*$                                              & 91709  & 91.7  &  87737  &  87.7   \\    
Prompt jets$^*$                                                   & 72075  & 78.6  &  84178  &  95.9   \\    
Prompt ${\Delta \phi (\mathrm{jet}_{1,2,3}, \vecptmiss)_{min}}^*$ & 49095  & 68.1  &  57261  &  68.    \\    
Prompt ${\Delta \phi (\mathrm{jet}_{j>3}, \vecptmiss)_{min}}^*$   & 27315  & 55.6  &  33832  &  59.1   \\    
Prompt ${\ptmiss / m_\mathrm{eff}(N_j)}^*$                        & 6670   & 24.4  &  18409  &  54.4   \\    
Prompt ${m_\mathrm{eff}\mathrm{(incl.)}}^*$                         & 6636   & 99.5  &  16848  &  91.5   \\   
DV jets                                                           & 6636   & 100.  &  16848  &  100.   \\
DV reconstruction$^\dag$                                          & 1524   & 23.   &  3850   &  22.9   \\    
DV fiducial                                                       & 1516   & 99.5  &  3825   &  99.4   \\    
DV material                                                       & 1494   & 98.5  &  3750   &  98.    \\  
$N_{\rm trk} \geq 2 $                                             & 1494   & 100.  &  3750   &  100.   \\       
$m_{\rm DV} > 5 $ GeV                                             & 88     & 5.9   &  265    &  7.1 \\
\hline 
\end{tabular*} 
\caption{\label{tab:cutflow_DGSATLASTuned}Numbers of simulated events
  $N$ and relative efficiencies $\epsilon$ (i.e.\ defined with respect
  to the previous cut) for our NMGMSB model with
  $c\tau_{\tilde{N}_{1}}=99$ mm (our P0 benchmark) at $\sqrt{s}=8$ TeV and $\sqrt{s}=13$
  TeV for our tuned cuts, as explained in the text. Events are generated 
  considering strong production. An asterisk denotes that the prompt cuts 
  are taken from signal regions
  {\tt 6jt-8} at $\sqrt{s}=8$ TeV and {\tt 4jt-13} at $\sqrt{s}=13$
  TeV as listed in Table~\protect\ref{tab:cuts}. The dagger is a reminder of the increased
  vertex merging distance of 5 
  mm.}
\end{table*} 

We used {\tt 6jt-8} for the prompt cuts at 8 TeV, however keeping in mind that the best sensitivity at 13 TeV is for the {\tt 4jt-13} signal region, we also perform efficiency calculations with the combination {\tt DVL} + {\tt 4jt-13}.  The efficiencies are shown in Table~\ref{tab:cutflow_DGSATLASTuned}.
The signal efficiency at 13 TeV is $\sim 0.2\%$. It would be desirable to relax the
prompt cuts further in order to increase this number, but a proper
estimate would require a full estimation of the DV background, which is
beyond the scope of this paper.  


However, an estimate of the contribution from heavy flavours may be
obtained. Given the hard multi-jet, $\ptmiss$ and $m_{\rm eff}$ cuts, the
dominant SM background is from $t \bar t + \mathrm{jets}$ production which is
also a source of $b$-hadrons and therefore a potential background for
DVs. In order to
examine this possibility, we simulate $10^6$ $t \bar t$ events,
and inspect the transverse impact parameter $d_0$ of the tracks coming
from the displaced $b$ vertices. We see that only a tiny fraction of
tracks pass $|d_0|>2$ mm from $t \bar t$ events ($\sim 1 \%$). Furthermore,
imposing the DV cuts (without any restrictions on hard jets), gives us an
efficiency of 0.1\% for $N_{\mathrm{trk}} \geq 2$ and imposing $m_{\mathrm{DV}} >
5$~GeV gives us no events at all.  We therefore do not expect any DV
contributions from heavy flavour once the hard jet cuts are made.   This
implies zero background events at 3.2 fb$^{-1}$ and we are already potentially
sensitive to signal cross sections of approximately 0.3 fb.  

The total strong sparticle production cross-section at 13 TeV before
cuts is 5.8~fb, and so with our illustrative cuts ({\tt DVL} + {\tt 4jt-13}), one would achieve
a signal cross-section after cuts of 0.01~fb.  With no expected background, 
the observation of a single event already corresponds to discovery, which for a gluino mass of $\sim2$~TeV 
(as in P0) is not achievable in prompt search channels with 100
fb$^{-1}$ at 13 TeV.  We may reasonably set the observation of at least three signal events as a requirement for discovery, which results in a best-case scenario of discovering a NMGMSB model with $m_{\tilde g} \sim 2$~TeV with 300 fb$^{-1}$ data at 13 TeV.  

We can also make an estimate of the worst-case scenario where there is a large DV background from systematic sources.  Such a background
occurs when a spurious track crosses an existing DV resulting in a reconstructed vertex satisfying the $N_\mathrm{trk}$ and $m_\mathrm{DV}$
requirements.  The ATLAS DV
analysis~\cite{Aad:2015rba} estimates only $\sim 0.4$ background vertices in
the DV + jet channel for the full 20 fb$^{-1}$ data of Run 
I (see Table 1 of
Ref.~\cite{Aad:2015rba}).  Given a $t \bar t$ production cross section $\sim
O(100~\mathrm{pb})$ at 8 TeV, this implies an efficiency $\sim 10^{-5}$.  To be
conservative about the effect of our relaxed cuts, we can assume that this
happens in about 1\% of events that pass the {\tt 4jt-13} cuts. Starting with
a total prompt background of $\sim 1$~fb (see Table 4 of
Ref.~\cite{Aaboud:2016zdn}) 
in the {\tt 4jt-13} channel as reported in the ATLAS analysis, we arrive at
0.01 fb for {\tt DVL} + {\tt 4jt-13}. A $3$-sigma discovery may then be viable
with $\sim$ 1 ab$^{-1}$ data at 13 TeV.  
 
This situation may be improved considerably by relaxing the prompt cuts.  An indication
 of where we may further relax the selection cuts comes from
examining the relative efficiencies at $8$ and $13$ TeV for the
cut on the ratio of $\ptmiss$ and $m_{\mathrm{eff}}(N_{j})$. We see that a change from $> 0.25$ at 8 TeV to $> 0.2$ at
13 TeV (see Table~\ref{tab:cuts}) already results in a gain of a factor 2.
Although, this is obviously also due to the increased energy of the overall
event, given that we have high $\ptmiss$ and $m_{\rm eff}$ cuts, an
additional factor of 3 may be gained by dropping the $\ptmiss / m_{\rm eff}(N_{j})$
cut altogether.

\begin{figure*}[ht]
\includegraphics[width=\textwidth]{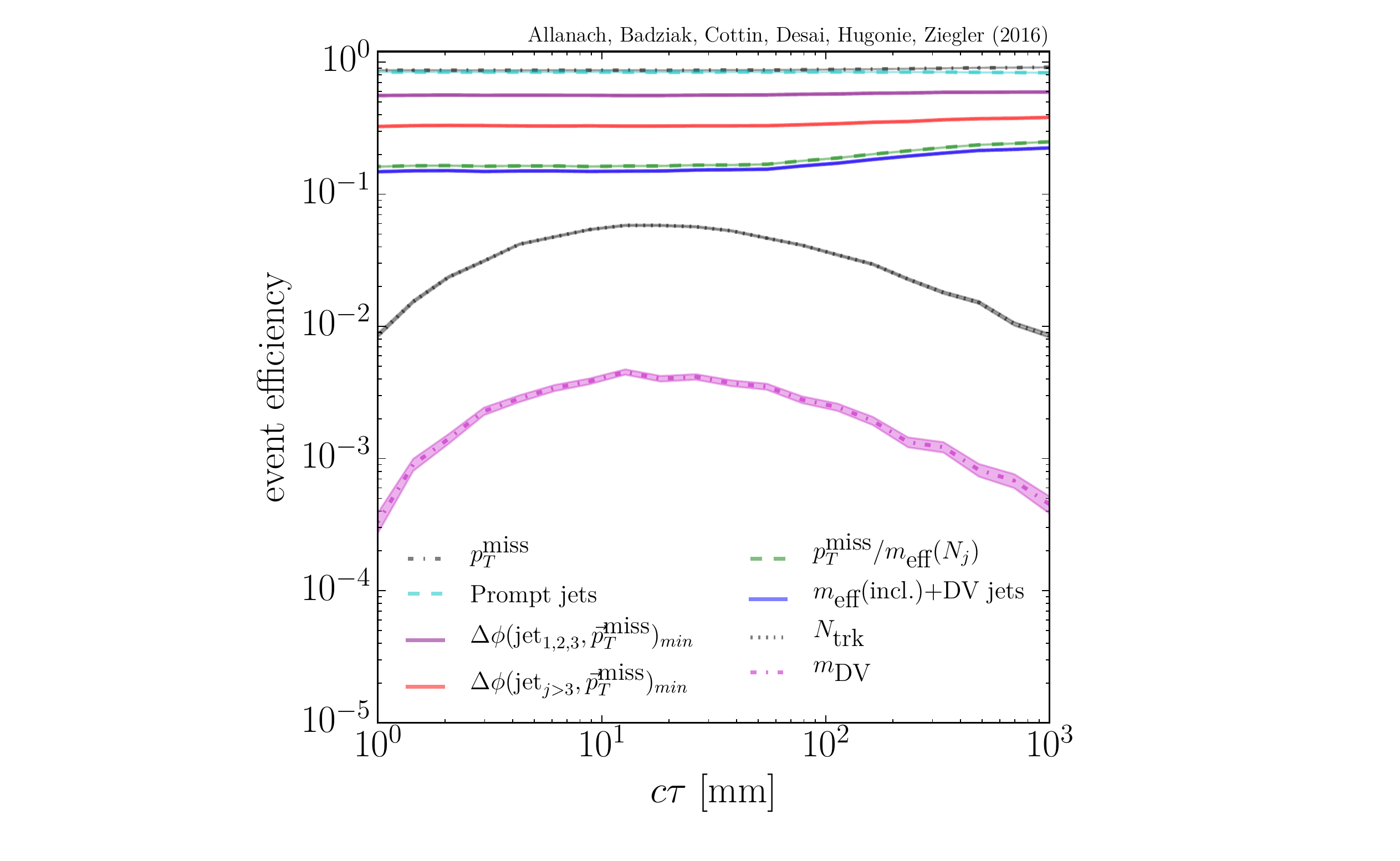}
\caption{Signal event efficiency of our simulation on a DGS benchmark
  with $m_{\tilde{g}}=1.96$ TeV, $m_{\tilde{N}_{1}}=98$ GeV,
  $m_{a_{1}}=23$ GeV (our P0 benchmark) against the lifetime of the long-lived singlino
  $c\tau_{\tilde{N}_{1}}$. Events are generated with $\sqrt{s}= 13 $
  TeV considering strong production. Independent prompt and DV cuts are
  presented. The singlino decay distance $c\tau$ has 
  been tweaked ``by hand'' in the SLHA files without changing soft parameters
  for the purposes of illustration.}
\label{effVsCtau_DGS13TeV}
\end{figure*}

We now study how the cut efficiencies behave with
singlino lifetime for benchmark P0. The result is shown in Figure~\ref{effVsCtau_DGS13TeV}, where we plot 
the effect of the cuts {\tt DVL} + {\tt 4jt-13} as a function
of the decay length $c\tau_{\tilde{N}_{1}}$. Note that we have merged the $m_{\mathrm{eff}}$(incl.) and the 
DV jets cuts together into one curve, as applying the DV jets cut after the $m_{\mathrm{eff}}$(incl.)
one does not change the number of events, for any lifetime (this can also be appreciated 
for P0 in Table \ref{tab:cutflow_DGSATLASTuned}).
We notice that standard prompt cuts are not very much affected by the singlino
lifetime, except for the cut on the ratio of
$p^{\mathrm{miss}}_T$ and $m_{\mathrm{eff}}(N_{j})$, which increases at higher
lifetimes. This is because $p^\mathrm{miss}_T$ is higher at high lifetimes, 
as explained in section~\ref{sec:currentBounds}. 


To summarise, with a combination of relaxed DV cuts and prompt SUSY
search cuts, one can discover a NMGMSB scenario with $m_{\tilde g} \sim
2$~TeV with 300 fb$^{-1}$ data which is not possible with prompt SUSY searches alone. 
With a full optimisation of relaxed DV cuts
+ prompt $\ptmiss$-based cuts, we may easily gain a further factor of ten in
the signal efficiency and given almost zero background, as shown above, one
could have higher sensitivity to the NMGMSB model in DV + prompt
searches as compared to prompt searches. We therefore strongly urge the
experiments to perform a dedicated background simulation with optimised
cuts.

\section{Summary \label{sec:summary}}

We have examined the prospects for discovery or
exclusion of the DGS-NMGMSB model. The model has some nice properties: the
SUSY flavour problem is addressed by gauge mediated SUSY breaking,
while the Higgs mass is made heavy enough through the mixing with the
NMSSM CP-even singlet. This singlet has a mass around 90 GeV, and therefore can be
made consistent with some small excesses in the LEP Higgs searches.
An interesting feature of the model is the presence of a gravitino LSP and 
a singlino-like neutralino NLSP that can be long-lived. 

As well as having the usual hard jets plus missing transverse momentum
signatures, the model predicts possible DVs from long-lived
singlinos.  These decay into $b \bar b$ and missing transverse
momentum in the form of gravitinos. However, displaced $b$'s are
somewhat problematic since $B$ mesons themselves travel a small
distance before visibly decaying and the `displaced displaced'
vertices have a very poor signal efficiency for getting past the
standard DV cuts. We have illustrated how loosening the
DV searches whilst imposing some prompt cuts to control
background results in significantly higher signal efficiency,
motivating a proper study with a full detector simulation (DV
analyses are difficult to perform accurately from outside the
experimental collaborations). We have provided a rough approximation to
the tracking efficiency that works for two General Gauge Mediation
models and one $R-$parity violating model over a range of possible
DV lifetimes, but clearly more work can be done to
provide a more comprehensive parameterisation.

We have re-cast current 8 TeV prompt searches to bound the gluino mass
from below at \runIbound{}, whereas current 13 TeV prompt searches are
less restrictive.  This is somewhat low compared to naive expectations
based on LHC exclusion results quoted for simplified models, but as
Figure~\ref{fig:specP0} shows, there are many different cascade decays
in the model. This means that the supersymmetric signal ends up being
shared out between many different channels, and may not be yet
detected in any single one~\cite{Aad:2015baa,Barr:2016inz}.  With
100~fb$^{-1}$ of integrated 
luminosity at 13 TeV though, the 0-lepton +  jets + $\ptmiss$
searches should be sensitive to up to \runIIbound. 

We further combine the search strategies in the prompt and displaced channels to demonstrate that a much better sensitivity could be obtained by optimising cuts.  In particular, we find that combining the relaxed DV cuts with the hard cuts from the 0-lepton +  jets + $\ptmiss$ analysis, a $> 3\sigma$ discovery can be made with 300 - 1000 fb$^{-1}$ data for a 2 TeV gluino mass depending on the systematic background. We indicate how this situation could be improved significantly by also relaxing some of the prompt cuts. It is clear that an optimised analysis in a DV + jets + $\ptmiss$ channel will yield better sensitivity than for either search method alone and we strongly urge the experimental collaborations to pursue this further.

\appendix
\section{Definitions of displaced observables \label{sec:def}}

\begin{figure*}[ht]
\centering
\includegraphics[width=\textwidth]{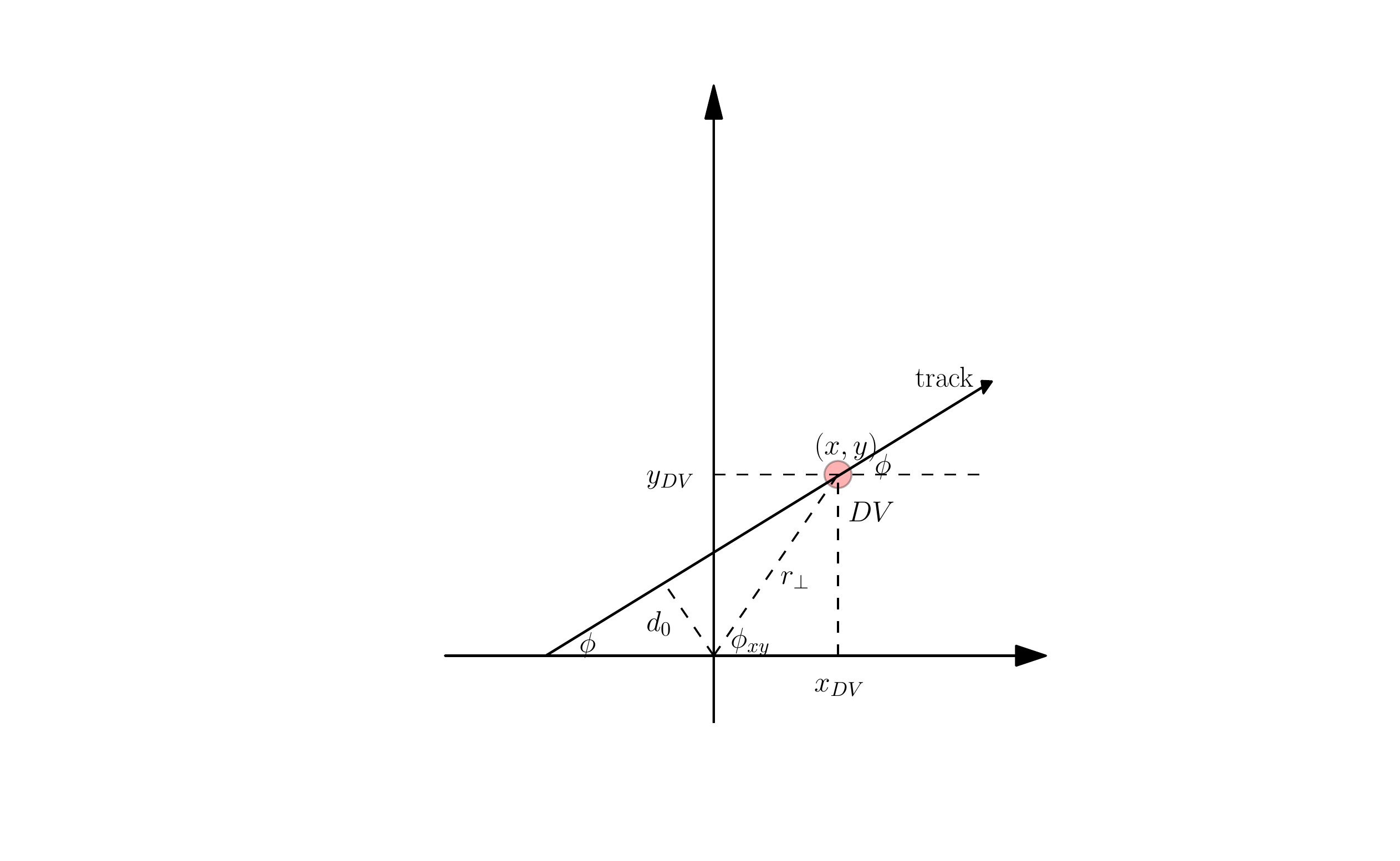}
\caption{Schematic view in the transverse $x-y$ plane of a displaced
  decay. The transverse impact parameter $d_{0}$ is defined with
  respect to the origin $(0,0,0)$. The daughter particle, which forms
  the track, was produced at $(x,y,z)$ from the the decay of a
  long-lived particle. The DV position is reconstructed
  from the average of all track's production vertices, represented by
  the pink disc.  }
\label{d0diagram}
\end{figure*}

Here we define the relevant observables used in the DV
re-cast. When a displaced track is produced at a point\footnote{The
  origin is defined to be the interaction point, and $z$ is along the
  beam line.} $(x,y,z)$, we define the transverse distance of the
truth track's production vertex to be
\begin{equation}
r_{\bot} = \sqrt{x^2+y^2}.
\end{equation}
Each track will have a transverse impact parameter $d_{0}$, which
corresponds to the distance of closest approach of the track to the
origin $(0,0,0)$ in the $x-y$ plane:
\begin{equation}
d_{0}=r_{\bot}\times\sin{(\phi_{xy}-\phi)},
\end{equation}
where $\phi$ is the azimuthal angle of the track, such that
$\tan{\phi}=p_{y}/p_{x}$ with $p_{x}$, $p_{y}$ the $x$ and $y$
component of the track momentum. $\phi_{xy}$ corresponds to the angle
in the transverse plane of the trajectory of the mother displaced
particle, as shown in Figure~\ref{d0diagram}.

Selected tracks are clustered together to form a DV.
The DV position $(x_{DV},y_{DV},z_{DV})$ is defined to
be the average position of all track's production points in that selected vertex.
The DV position in the transverse plane is
defined to be
\begin{equation}
r_{DV} = \sqrt{x^2_{DV}+y^2_{DV}}. 
\end{equation}
The DV position $(x_{DV},y_{DV},z_{DV})$ is equal to the
truth decay position of the mother particle in principle, but in our
simulation it is defined from the displaced tracks, using the truth
information for the production positions, as explained above.

\section*{Acknowledgements}
The authors acknowledge the support of France Grilles for providing
cloud computing resources on the French National Grid Infrastructure.
This work has been partially supported by STFC grant ST/L000385/1, by
the the Office of High Energy Physics of the U.S. Department of Energy
under Contract DE-AC02-05CH11231, by the National Science Foundation
under grant PHY-1316783, by the Foundation for Polish Science through
its programme HOMING PLUS, by National Science Centre under
research grant DEC-2014/15/B/ST2/\linebreak 02157, by the  ILP  LABEX under reference ANR-10-LABX-63, and by
French state funds managed by the ANR within the Investissements d'Avenir programme under reference ANR-11-IDEX-0004-02. GC was funded by the
postgraduate Conicyt-Chile Cambridge Scholarship 84130011.  ND was
partially supported by the Alexander von Humboldt Foundation.  MB
acknowledges support from the Polish Ministry of Science and Higher
Education (decision no.\ 1266/MOB/IV/2015/0).  BCA, MB and GC would
like to thank other members of the Cambridge SUSY Working group for
discussions. RZ thanks B. Fuks, A. Mariotti, D. Redigolo and O. Slone for useful discussions. We thank the authors of {\tt Delphes3} for discussions
about the program related to this work.  ND and MB would like to thank
the Cavendish Laboratory for hospitality offered while working on this
project.  MB and RZ thanks the Galileo Galilei Institute for
Theoretical Physics and INFN for hospitality and partial support
during the completion of this work.

\bibliographystyle{JHEP-2}
\bibliography{dgsPheno}

\end{document}